\documentclass[12pt,epsf]{article}
\usepackage{graphicx}
\begin{document}
\def\slashchar#1{\setbox0=\hbox{$#1$} 
\dimen0=\wd0 
\setbox1=\hbox{/} \dimen1=\wd1 
\ifdim\dimen0>\dimen1 
\rlap{\hbox to \dimen0{\hfil/\hfil}} 
#1 
\else 
\rlap{\hbox to \dimen1{\hfil$#1$\hfil}} 
/ 
\fi}

\def\a{\alpha}
\def\b{\beta}
\def\c{\varepsilon}
\def\d{\delta}
\def\e{\epsilon}
\def\f{\phi}
\def\g{\gamma}
\def\h{\theta}
\def\k{\kappa}
\def\l{\lambda}
\def\m{\mu}
\def\n{\nu}
\def\p{\psi}
\def\q{\partial}
\def\r{\rho}
\def\s{\sigma}
\def\t{\tau}
\def\u{\upsilon}
\def\v{\varphi}
\def\w{\omega}
\def\x{\xi}
\def\y{\eta}
\def\z{\zeta}
\def\D{\Delta}
\def\G{\Gamma}
\def\H{\Theta}
\def\L{\Lambda}
\def\F{\Phi}
\def\P{\Psi}
\def\S{\Sigma}

\def\o{\over}
\def\beq{\begin{eqnarray}}
\def\eeq{\end{eqnarray}}
\newcommand{\gsim}{ \mathop{}_{\textstyle \sim}^{\textstyle >} }
\newcommand{\lsim}{ \mathop{}_{\textstyle \sim}^{\textstyle <} }
\newcommand{\vev}[1]{ \left\langle {#1} \right\rangle }
\newcommand{\bra}[1]{ \langle {#1} | }
\newcommand{\ket}[1]{ | {#1} \rangle }
\newcommand{\EV}{ {\rm eV} }
\newcommand{\KEV}{ {\rm keV} }
\newcommand{\MEV}{ {\rm MeV} }
\newcommand{\GEV}{ {\rm GeV} }
\newcommand{\TEV}{ {\rm TeV} }
\def\diag{\mathop{\rm diag}\nolimits}
\def\Spin{\mathop{\rm Spin}}
\def\SO{\mathop{\rm SO}}
\def\O{\mathop{\rm O}}
\def\SU{\mathop{\rm SU}}
\def\U{\mathop{\rm U}}
\def\Sp{\mathop{\rm Sp}}
\def\SL{\mathop{\rm SL}}
\def\tr{\mathop{\rm tr}}

\def\IJMP{Int.~J.~Mod.~Phys. }
\def\MPL{Mod.~Phys.~Lett. }
\def\NP{Nucl.~Phys. }
\def\PL{Phys.~Lett. }
\def\PR{Phys.~Rev. }
\def\PRL{Phys.~Rev.~Lett. }
\def\PTP{Prog.~Theor.~Phys. }
\def\ZP{Z.~Phys. }


\baselineskip 0.7cm

\begin{titlepage}

\begin{flushright}
IPMU-10-0222\\
UT-10-22
\end{flushright}

\vskip 1.35cm
\begin{center}
{\large \bf 
Gravitino Dark Matter and Light Gluino in an R-invariant \\ Low Scale Gauge Mediation
}
\vskip 1.2cm
Masahiro Ibe, Ryosuke Sato, Tsutomu T. Yanagida and Kazuya Yonekura
\vskip 0.4cm

{\it Institute for the Physics and Mathematics of the Universe, University of Tokyo, \\
Kashiwa 277-8568, Japan}\\
{\it  
Department of Physics, University of Tokyo, \\ 
Tokyo 113-0033, Japan }

\vskip 1.5cm

\abstract{ 
We consider the simplest class of the R-invariant gauge mediation model
with the gravitino mass in the one to ten keV range.  
We show that the entropy production from the supersymmetry breaking sector
makes the gravitino into a warm dark matter candidate.
We also discuss that the gluino mass can be lighter than the wino
mass even when the messenger sector satisfies the GUT relations
at the GUT scale. 
}
\end{center}
\end{titlepage}

\setcounter{page}{2}

\section{Introduction}
Dark matter is an important clue to the theory beyond the standard model (SM).
There have been proposed a lot of models to explain dark matter. 
The light gravitino is a very interesting candidate for dark matter among them, 
since the gravitino itself is a unique and inevitable prediction of supergravity (SUGRA).
If the gravitinos were in the thermal equilibrium in the early universe, 
the gravitino mass $m_{3/2}$ is required to be $m_{3/2}\simeq 100$\,eV from the 
observed dark matter density, $\Omega_{\rm DM} \simeq 0.1$. 
This prediction is very much interesting, since we can test the gravitino dark matter hypothesis at LHC. 
The gravitino mass, $m_{3/2}\simeq 100$\,eV, is, however,  too small 
to be the cold dark matter and is disfavored 
for the successful galaxy formation\,\cite{Viel:2005qj}. 

The above argument is based on an unjustified assumption on the thermal history of the early universe. 
In fact, if we had late time entropy production after the decoupling time of the gravitino, 
the mass of the gravitino dark matter may be raised up to a few keV.
Moreover, the gravitino dark matter with a mass in the one to ten keV range serves 
as the warm dark matter which has recently been invoked 
as possible solutions to the seeming discrepancies 
between the  observation and the simulated results of the galaxy formation
based on the cold dark matter scenario\,\cite{WDM}.%
\footnote{
The detailed analyses in Refs.\,\cite{Boyarsky:2008xj,Maccio':2009rx}
have placed lower bounds on the warm dark matter mass around a few keV range.
Thus, it is safer to assume that the dark matter is in the ten keV range.
}

In this paper, we discuss the late-time entropy production
from the SUSY breaking sector. 
As we will show,  large entropy can be produced from the SUSY breaking sector
when the sector has meta-stable particles whose lifetimes
are long enough to dominate the energy density of the universe before they decay.
As a result, the gravitino with a mass in the one to ten keV range
can be a good candidate for the warm dark matter 
with the help of the entropy production.

The gravitino mass in a range of the one to ten keV also 
has an interesting implications on the phenomenological aspects of
the supersymmetric standard model (SSM).
For the gravitino mass in the keV range, we are led to consider 
the models with gauge mediation~\cite{Dine:1981za,mGMSB} where the SUSY breaking effects
are mediated to the SSM sector via the gauge interactions.

In this paper, as a particular example of the models with gauge mediation,
we consider a class of the direct mediation models developed in Ref.\,\cite{Izawa:1997gs},
where the SUSY breaking vacuum is stable. 
The important feature of this class of models is that the models possess 
an R-symmetry.
It should be noted that, independent of the SUSY-breaking mediation scheme,   
the (discrete) R-symmetry is considered to be a crucial symmetry for any low-energy 
SUSY extension of the Standard Model.
This can be seen from the fact that SUSY should be broken at very high energy scale 
to obtain the nearly vanishing cosmological constant if the R-symmetry is largely broken 
by the constant term of the superpotential in supergravity. 
Therefore, it is quite tempting to consider a mediation mechanism which possesses an R-symmetry.

The notable feature of this class of gauge mediation models is 
a peculiar spectrum of the superparticles.
Especially, the gaugino masses do not satisfy 
the so-called the Grand Unified Theory (GUT) relations
even if the masses and the couplings 
of the messenger fields satisfying the GUT relations at the GUT scale~\cite{Nomura:1997uu}.
For example, the light gluino of mass 300\,GeV$-$1\,TeV is achieved with
the heavier wino of mass 500\,GeV$-$ 2 TeV even 
for the boundary condition satisfying the GUT relations at the GUT scale.
Such a light gluino will be easily produced at the LHC,
and hence, almost all the parameter space is expected to be probed by the LHC experiment.

The organization of the paper is as follows.
In section\,\ref{sec:RinvGM}, 
we discuss the mass spectrum of the SSM particles in
the simplest class of the R-invariant gauge mediation model for the gravitino with a mass 
in the one to ten keV range.
There, we show that the typical gaugino mass spectrum is distinctive
from the so-called minimal gauge mediation model.
In section\,\ref{sec:Entropy}, we discuss an entropy
production mechanism which makes the gravitino dark matter scenario with a mass
in the one to ten keV range consistent with the observed dark matter density.
The final section is devoted to our conclusions.

\section{An R-invariant gauge mediation model}\label{sec:RinvGM}
Let us discuss the minimal R-invariant gauge mediation model developed in Ref.\,\cite{Izawa:1997gs}. 
We introduce two pairs of massive messengers, $\Psi_{i}, {\tilde \Psi}_{i}$ 
with $i=1,2$. Here, $\Psi_{i}$ and ${\tilde \Psi}_{i}$ transform as ${\bf 5}$ and ${\bf 5^*}$ 
in terms of the minimal $SU(5)$ GUT representations, respectively. 
We further introduce a SUSY-breaking gauge singlet field $S$ which 
has non-vanishing expectation values of the $F$ and $A$ terms,
\begin{equation}
\vev{S(x,\theta)}=\vev S + F\theta^2.
\end{equation}
(We abuse the notation for chiral fields and its lowest components.)
We assume, throughout this paper, that the $F$ term is the dominant component of the SUSY breaking and hence
the gravitino mass is given by
\begin{equation}
m_{3/2}\simeq \frac{|F|}{\sqrt{3}M_P}.
\end{equation}
Here, $M_P\simeq 2.4\times 10^{18}$ GeV is the reduced Planck mass. 

Let us assume that the superpotential in the messenger sector is given by,
\begin{equation}
\label{eq:messenger}
W=\left( \tilde{\Psi}_1, \tilde{\Psi}_2 \right) \left(
\begin{array}{cc}
k S & m \\
m & 0 
\end{array}
\right)
\left(
\begin{array}{c}
\Psi_1 \\
 \Psi_2
\end{array}
\right)\ ,
\end{equation}
where $k$ and $m$ denote the coupling constant and the mass parameter, respectively.
We see that the above superpotential is invariant under an R-symmetry
with the charge assignment, $S(2),~\Psi_1(0),~\tilde{\Psi}_1(0),~\Psi_2(2),~\tilde{\Psi}_2(2)$.
Notice that the vacuum expectation value of the
scalar component of $S$ breaks the R-symmetry spontaneously. 

So far, we have treated messenger fields in an $SU(5)_{\rm GUT}$ symmetric way.
Below the GUT scale, however, 
the $SU(5)_{\rm GUT}$ messenger multiplets split into 
$\Psi_i \to (\Psi^{(d)}_{i},\Psi^{(\ell)}_{i})$ and $\tilde{\Psi}_i \to (\tilde{\Psi}^{(d)}_{i},\tilde{\Psi}^{(\ell)}_{i})$ which transform as $(\bf{3}_{-1/3},\bf{2}_{1/2})$ and $(\bf{3}^*_{1/3},\bf{2}_{-1/2})$ 
under the SM gauge groups, $SU(3)_C \times SU(2)_L \times U(1)_Y$, respectively. 
In the followings, we name the messengers with the superscripts $d$ and $\ell$ 
``down-type'' and ``lepton-type'', respectively.
In accordance with the above splitting, the coupling constants 
and the mass parameters in Eq.\,(\ref{eq:messenger}) 
may take different values for each type of messengers, i.e. $k^{(d,\ell)}$ and $m^{(d,\ell)}$.
Especially, the renormalization group (RG) evolution makes them different at the 
lower energy scale, even if we impose $k^{(d)}=k^{(\ell)} $ and $m^{(d)}=m^{(\ell)} $ at the GUT scale. 

In this model, 
we can take $k^{(\chi)}$, $m^{(\chi)}$, $\langle S\rangle$ and $F$ as real positive by the phase rotation of the fields without loss of generality.
Therefore, we can avoid $CP$ violation in the present model.
In the following of this paper, we take these parameters as real positive.

Let us discuss the condition for the messenger scalar not to be tachyonic.
The mass parameters are required to satisfy $k^{(d)}F/m^{(d)2} < 1$ and $k^{(\ell)}F/m^{(\ell)2} < 1$
for the messenger fields not to be tachyonic. 
By the RG effect, the condition for the lepton-type messenger gives severer constraints than the one for the down-type messenger.
This fact can be seen as follows. The ratio of $k^{(d)}F/m^{(d)2}$ and $k^{(\ell)}F/m^{(\ell)2}$ evolves according the RG equations,
\beq
\frac{d}{d\log \m} \log \left( \frac{k^{(d)} F}{m^{(d)2}} \bigg/ \frac{k^{(\ell)} F}{m^{(\ell)2}} \right)
&=& -2\left(\gamma_{\Psi^{(d)}_2}-\gamma_{\Psi^{(\ell)}_2}\right) \nonumber\\
&\simeq& \frac{16}{3}\frac{\alpha_3}{4\pi}-3\frac{\alpha_2}{4\pi}-\frac{1}{3}\frac{\alpha_1}{4\pi},
\eeq
where we have used the RG equations of $k$ and $m$
in terms of  the anomalous dimensions of $\Psi^{(\chi)}_i(\tilde{\Psi}^{(\chi)}_i)$ and $S$, 
 $\gamma_{\Psi^{(\chi)}_i}$ and $\gamma_S$,
\beq
\frac{\partial}{\partial \log \mu}k^{(\chi)} &=& (2\gamma_{\Psi^{(\chi)}_1}+\gamma_S)k^{(\chi)}, \\
\frac{\partial}{\partial \log \mu}m^{(\chi)} &=& (\gamma_{\Psi^{(\chi)}_1}+\gamma_{\Psi^{(\chi)}_2})m^{(\chi)},~~~~~~~~(\chi=d,\ell)\ .
\eeq
If we require $k^{(d)} = k^{(\ell)}$ and $m^{(d)} = m^{(\ell)}$ at the GUT scale, 
we can get 
\beq
\frac{k^{(d)} F}{m^{(d)2}} \bigg/ \frac{k^{(\ell)} F}{m^{(\ell)2}}
&\simeq& \exp\left[ -\int_{M_{\rm med}}^{M_{\rm GUT}} d\log\m \left(\frac{16}{3}\frac{\alpha_3}{4\pi}-3\frac{\alpha_2}{4\pi}-\frac{1}{3}\frac{\alpha_1}{4\pi}\right) \right] \label{eq:stablityratio}
\eeq
at the mediation scale.
Due to the strong $SU(3)_C$ effect, $k^{(d)} F/m^{(d)2}$ becomes smaller than
$k^{(\ell)} F/m^{(\ell)2}$ at the mediation scale. 
Therefore, the condition $k^{(\ell)} F/m^{(\ell)2} < 1$ guarantees that all the messenger scalars have positive squared masses.
The fact $k^{(d)} F/m^{(d)2}<k^{(\ell)} F/m^{(\ell)2}$ has an interesting consequence for the gaugino masses as will see later.

\subsection*{Mass spectrum of the present model}
The gaugino masses are given by
\beq
M_1 &=& \frac{\alpha_1}{2\pi}\left(\frac{2}{5}\L^{(d)}_{1/2}+\frac{3}{5}\L^{(\ell)}_{1/2} \right)\ , \label{eq:binomass}\\
M_2 &=& \frac{\alpha_2}{2\pi}\L^{(\ell)}_{1/2}\ , \label{eq:winomass}\\
M_3 &=& \frac{\alpha_3}{2\pi}\L^{(d)}_{1/2}\ , \label{eq:gluinomass}
\eeq
and the squared masses of sfermion $\tilde f$ are given by
\beq
m^2_{\tilde f}=2\left(\frac{\alpha_1}{4\pi} \right)^2C_1\left(\frac{2}{5}\L^{(d)2}_0+\frac{3}{5}\L^{(\ell)2}_0\right)+2\left(\frac{\alpha_2}{4\pi} \right)^2C_2\,\L^{(\ell)2}_0
+2\left(\frac{\alpha_3}{4\pi} \right)^2C_3\,\L^{(d)2}_0,
\eeq
where $\alpha_a~(a=1,2,3)$  are gauge coupling fine structure constants of $U(1)_Y$, $SU(2)_L$, $SU(3)_C$, and
$C_a~(a=1,2,3)$ are quadratic casimir invariants%
\footnote{We use the GUT normalization for the $U(1)_Y$ gauge group. In particular,
the quadratic casimir is given by $C_1=\frac{3}{5}Y^2$ in terms of the hypercharge $Y$.} 
of the sfermion $\tilde f$ under the group $U(1)_Y$, $SU(2)_L$, $SU(3)_C$.
Here, $\L^{(\chi)}_{1/2}$ and $\L^{(\chi)2}_0$ ($\chi=d,\ell$) are functions of $m^{(\chi)}, k^{(\chi)}\vev{S}$ and $k^{(\chi)}F$
whose explicit forms can be read from Ref.~\cite{Nomura:1997uu,Sato:2009dk}.

\subsubsection*{Gaugino-sfermion mass ratio}
In the case of the so-called minimal gauge mediation (mGM)~\cite{mGMSB}, with a messenger superpotential of the form $(m^{(\chi)}+k^{(\chi)}S)\tilde{\Psi}\Psi$,
both the $\L^{(\chi)}_{1/2}$ and $\L^{(\chi)}_{0}$ are of order 
\beq
\L^{(\chi)}_{1/2}|_{\rm mGM} \sim \L^{(\chi)}_{0}|_{\rm mGM} \sim \frac{k^{(\chi)}F}{m^{(\chi)}}\ ,
\eeq
where we have neglected $\vev{S}$ (inclusion of it is straightforward).
However, there is a significant difference in the gaugino masses in the present model.
The rough behavior of the soft masses are as follows.
For $k^{(\chi)}\vev{S} \lsim m^{(\chi)}$,
the soft masses $\L^{(\chi)}_{1/2}$ and $\L^{(\chi)2}_0$ are approximately given by
\beq
\L^{(\chi)}_{1/2} &\sim &  {\cal O}(0.1) \times \frac{(k^{(\chi)}\vev{S})(k^{(\chi)}F)^3}{|m^{(\chi)}|^6}\left(1+{\cal O}\left(\left|\frac{k^{(\chi)}F}{m^{(\chi)2}}\right|^2\right)\right), \label{eq:gauginoLambda} \\
\L^{(\chi)2}_0 &\sim& {\cal O}(1) \times\left( \frac{k^{(\chi)}F}{m^{(\chi)}}\right)^2 \left(1+{\cal O}\left(\left|\frac{k^{(\chi)}F}{m^{(\chi)2}}\right|^2 \right)\right) ,
\eeq
and they decrease when $k^{(\chi)}{S}$ becomes much larger than $m^{(\chi)}$.
Notice that there are no terms of order $k^{(\chi)}F/m^{(\chi)}$ in the gaugino masses, which would be present in the minimal gauge mediation.
Because the mass parameters are required to satisfy $k^{(\chi)}F < m^{(\chi)2}$
for the messenger fields not to be tachyonic,
the above approximated expressions show that the gaugino masses are suppressed 
compared with the sfermion masses. 

\subsubsection*{Wino-gluino mass ratio}
In the case of the minimal gauge mediation, the ratio of gaugino mass contributions from the down-type and the lepton-type messengers, 
$\L^{(d)}_{1/2}|_{\rm mGM} $ and  $\L^{(\ell)}_{1/2}|_{\rm mGM}$, is given by
\beq
\left. \frac{\L^{(d)}_{1/2} }{\L^{(\ell)}_{1/2}} \right|_{\rm mGM} \simeq \left( \frac{k^{(d)}}{k^{(\ell)}} \right) \cdot \left( \frac{m^{(d)}}{m^{(\ell)}} \right)^{-1}
\eeq
where we have neglected the higher order terms in $k^{(\chi)}F/m^{(\chi)2}$. One can easily check that this ratio is invariant under the RG flow%
\footnote{
As pointed out in Ref.~\cite{Shirai:2010rr}, this RG invariance is a general feature of models where gaugino masses are generated at the leading order
in the SUSY breaking $F$-term. In the effective field theory language, the leading term in $F$ is generated~\cite{Giudice:1997ni} by a holomorphic 
gauge kinetic function
$\int d^2\theta h(S)W^\a W_\a$ and this holomorphic term is independent of the wave function renormalization of the messenger fields.}.
Thus, if we impose the GUT relations, $k^{(d)}=k^{(\ell)}$ and $m^{(d)}=m^{(\ell)}$, at the GUT scale, we have $\L^{(d)}_{1/2} \simeq \L^{(\ell)}_{1/2}$
at the messenger scale. This indicates that the gaugino masses obey the famous GUT relation $M_{\rm bino} : M_{\rm wino} : M_{\rm gluino}\simeq \alpha_1:\alpha_2:\alpha_3 \simeq 1:2:6$.

In the present model, on the other hand, the ratio 
$\L^{(d)}_{1/2}/\L^{(\ell)}_{1/2}$ is given by
\beq
\frac{\L^{(d)}_{1/2} }{\L^{(\ell)}_{1/2}} \simeq \left( \frac{k^{(d)}}{k^{(\ell)}} \right)^4 \cdot \left( \frac{m^{(d)}}{m^{(\ell)}} \right)^{-6}. \label{eq:gauginoratio}
\eeq
%
By using similar argument to derive Eq. (\ref{eq:stablityratio}),
the ratio Eq.~(\ref{eq:gauginoratio}) at the scale of the gauge mediation, $M_{\rm med}={\cal O}(m^{(d,\ell)})$, is roughly given by 
\beq
\frac{\L^{(d)}_{1/2} }{\L^{(\ell)}_{1/2}} &\simeq & \exp\left[-\int^{M_{\rm GUT}}_{M_{\rm med}} d \log \mu \left(2(\gamma_{\Psi^{(d)}_1}-\gamma_{\Psi^{(\ell)}_1})
-6(\gamma_{\Psi^{(d)}_2}-\gamma_{\Psi^{(\ell)}_2})\right) \right]
 \nonumber \\
&\simeq&
\exp \left[-\int^{M_{\rm GUT}}_{M_{\rm med}} d \log \mu \left(\frac{32}{3}\frac{\alpha_3}{4\pi}-6\frac{\alpha_2}{4\pi}-\frac{2}{3}\frac{\alpha_1}{4\pi} +\frac{1}{2\pi}\frac{k^{(d)2}-k^{(\ell)2}}{4\pi} \right) \right] \label{eq:gauginoratio2},
\eeq
if we impose the GUT relations at the GUT scale, $M_{\rm GUT}$, i.e. 
$m^{(d)}=m^{(\ell)}$ and $k^{(d)}=k^{(\ell)}$. 
The strong $SU(3)_C$ interaction makes $\L^{(d)}_{1/2}$ smaller than
$\L^{(\ell)}_{1/2}$ at the mediation scale. 
Therefore, from Eqs.\,\,(\ref{eq:binomass})--(\ref{eq:gluinomass}), 
we see that the gluino mass is rather suppressed~\cite{Nomura:1997uu} in the present model compared with 
the one in the minimal gauge mediation. 
Furthermore, when $k^{(\ell)}F \simeq m^{(\ell)2}$, the ratio $\L^{(d)}_{1/2} / \L^{(\ell)}_{1/2}$ is more suppressed than Eq.~(\ref{eq:gauginoratio2}).
This is because the higher order contributions of $k^{(\ell)}F/m^{(\ell)2}$ in Eq.~(\ref{eq:gauginoLambda}) make $\L^{(\ell)}_{1/2}$ enhanced,
while $\L^{(d)}_{1/2}$ is not so enhanced since the RG equation indicates $k^{(d)}F/m^{(d)2} < k^{(\ell)}F/m^{(\ell)2}$ as we have discussed above.

\subsection*{Numerical results}
\begin{figure}[tbp]
\begin{center}
\begin{minipage}{.46\linewidth}
\includegraphics[width=\linewidth]{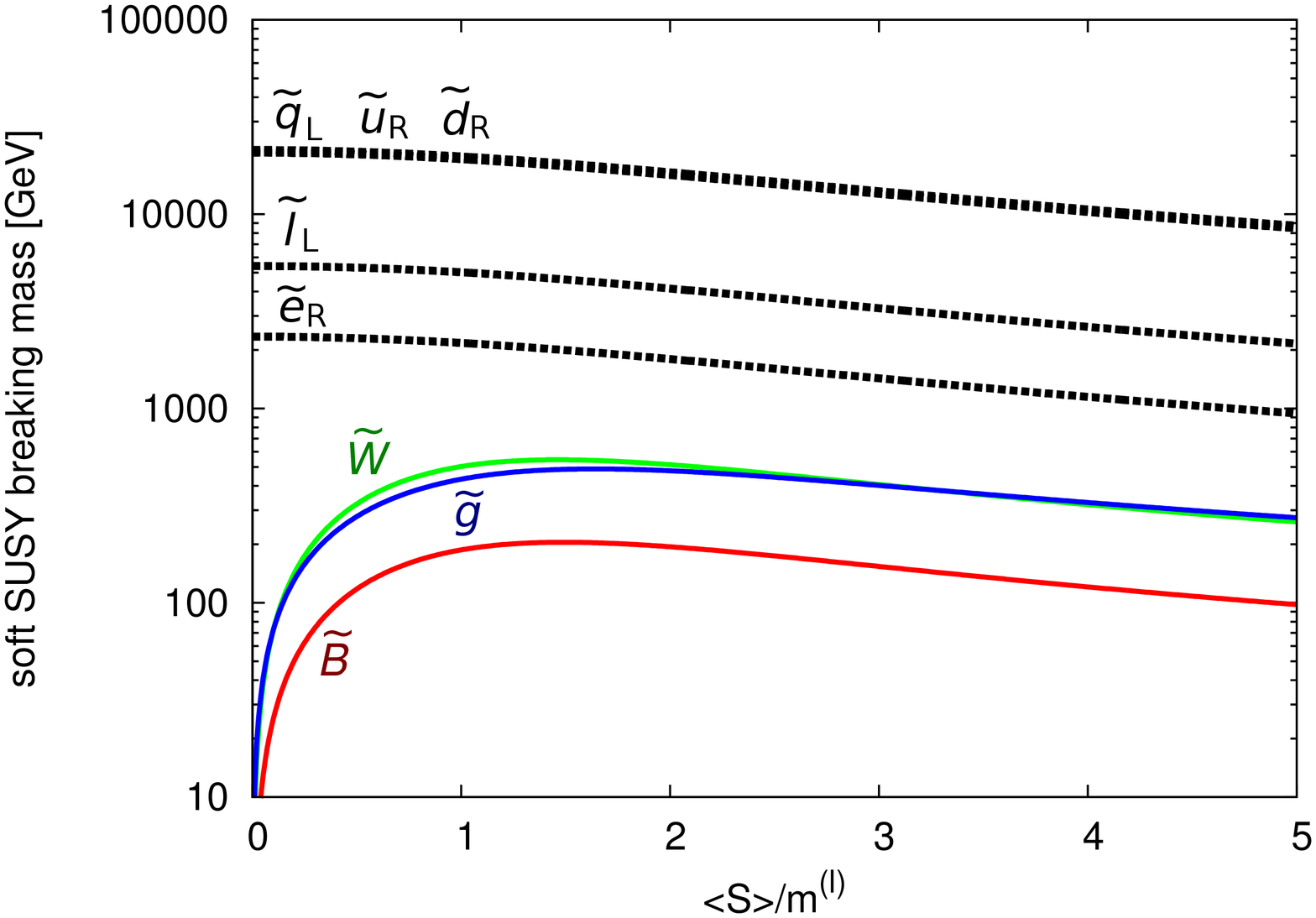}
\end{minipage}
\hspace{.02\linewidth}
\begin{minipage}{.46\linewidth}
\includegraphics[width=\linewidth]{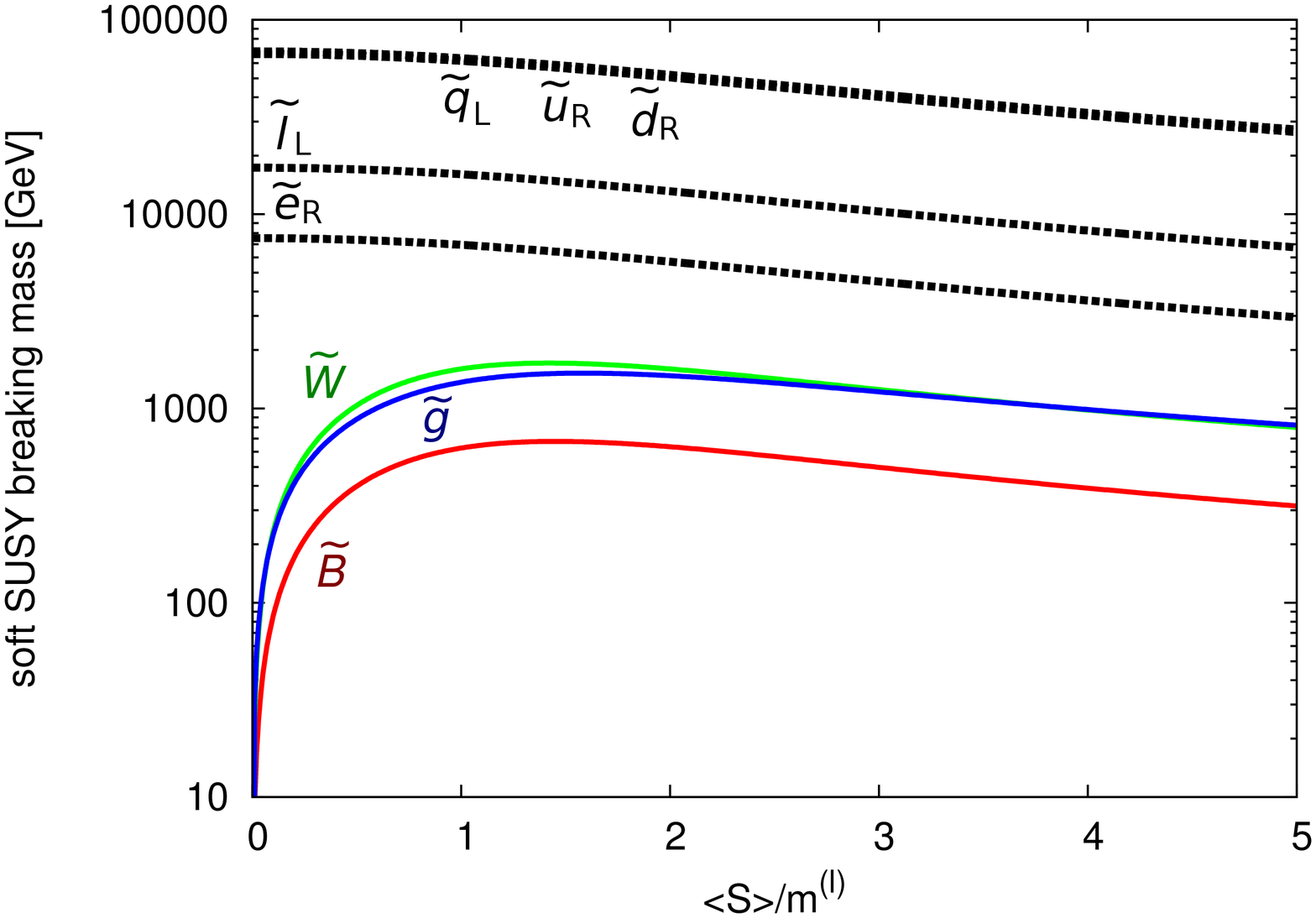}
\end{minipage}
\caption{
The soft SUSY breaking masses of gauginos and sfermions in the minimal R-invariant model
for $m_{3/2}=1$\,keV (left) and $m_{3/2}=10$\,keV (right).
We set $k^{(\ell)} F/m^{(\ell)2}=0.9$.
}\label{fig:1kev_z}
\vspace{2cm}
\begin{minipage}{.46\linewidth}
\includegraphics[width=\linewidth]{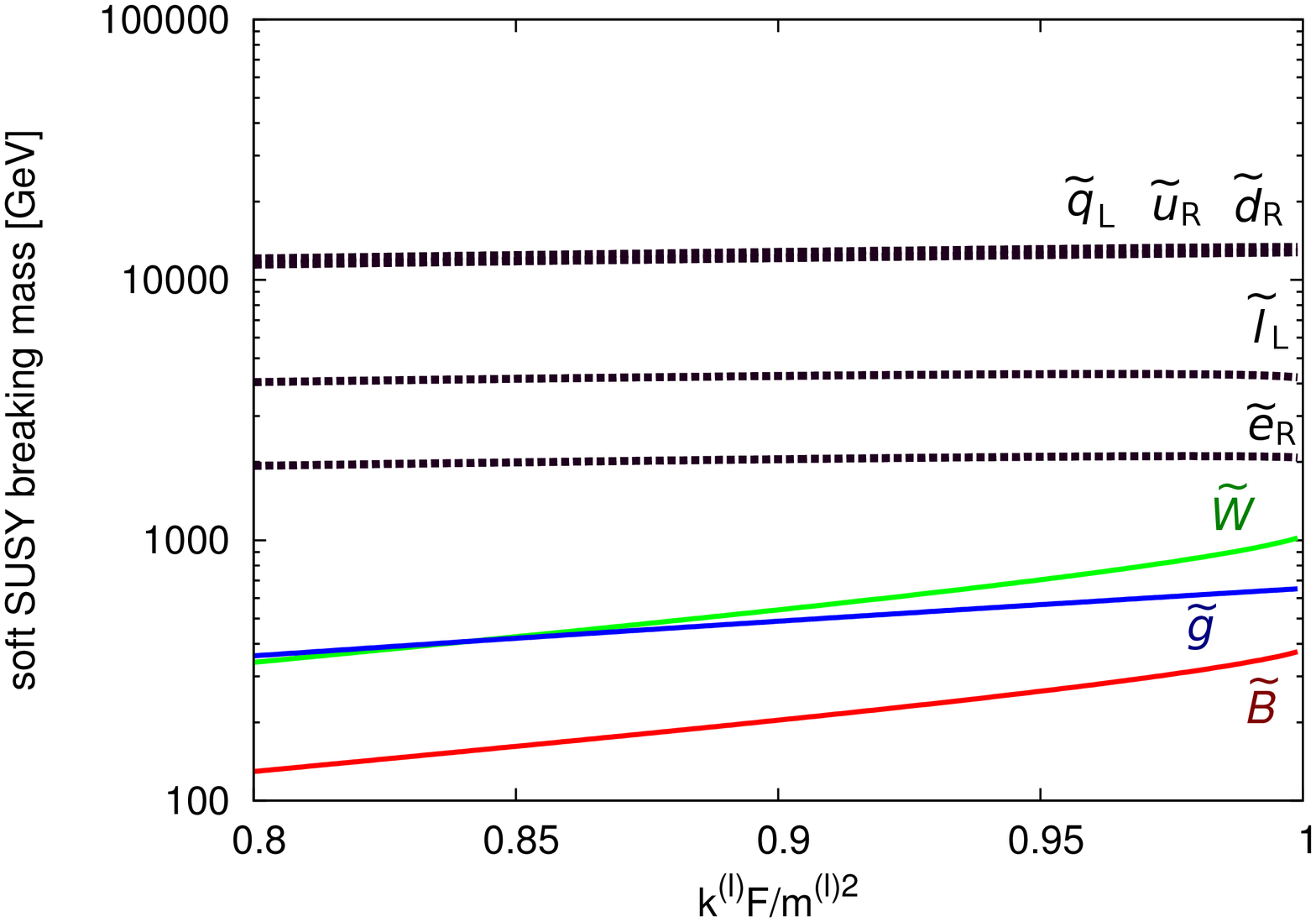}
\end{minipage}
\hspace{.02\linewidth}
\begin{minipage}{.46\linewidth}
\includegraphics[width=\linewidth]{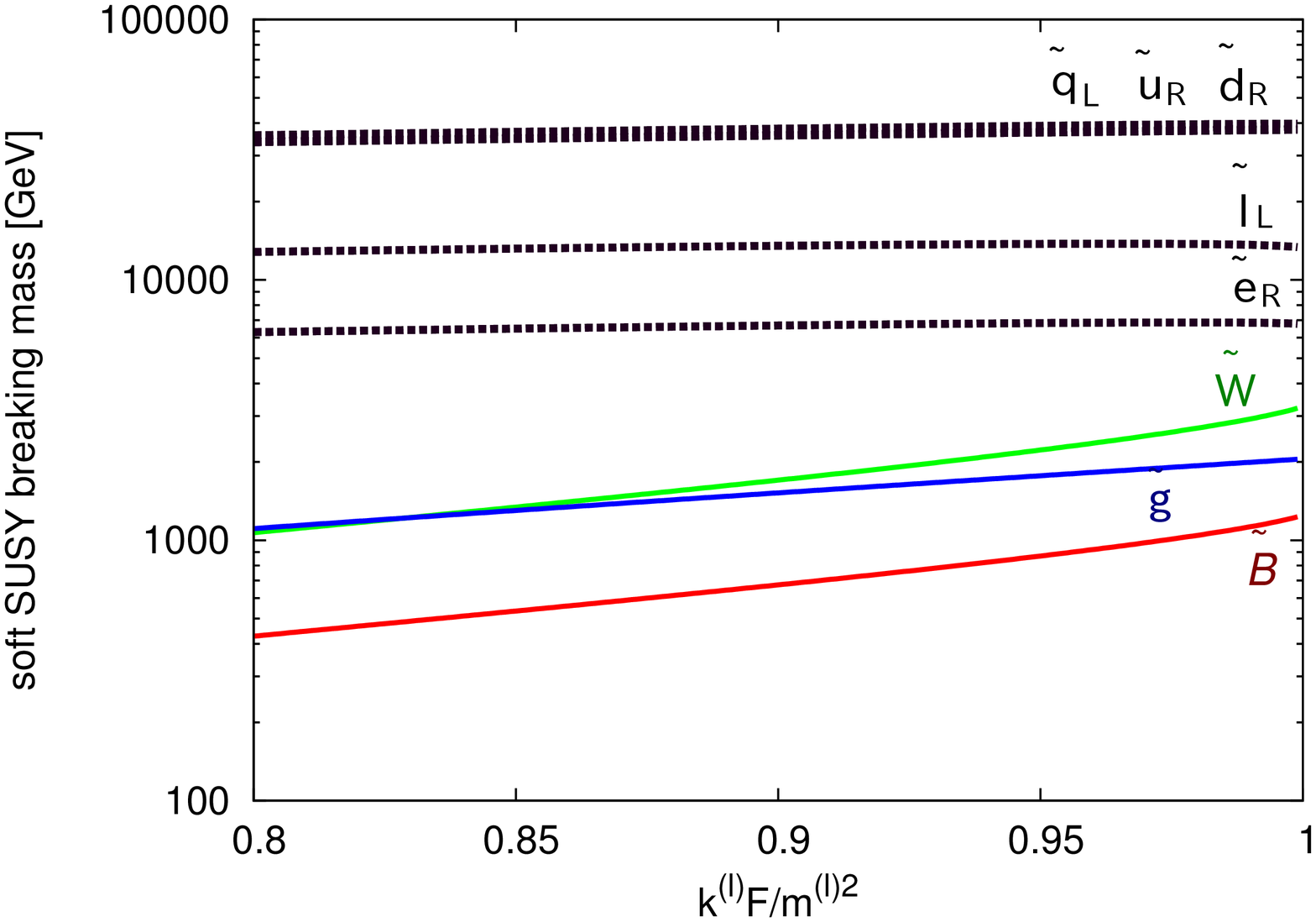}
\end{minipage}
\caption{
The soft SUSY breaking masses of gaugino and sfermion in the minimal R-invariant model 
for $m_{3/2}=1$\,keV (left) and $m_{3/2}=10$\,keV (right).
We take $\langle S\rangle$ as a value which maximizes the gluino mass.
}\label{fig:1kev}
\end{center}
\end{figure}
We show numerical results of the soft SUSY breaking masses of gauginos and sfermions 
in the minimal R-invariant gauge mediation model.
We impose $m^{(d)}=m^{(\ell)}$ and $k^{(d)}=k^{(\ell)}$ at the GUT scale.
The larger soft masses are obtained for the larger coupling constants, $k^{(d)}$ and $k^{(\ell)}$
at the mediation scale.
The coupling constants are, however, 
not able to be arbitrary large, because the Yukawa-type interactions are not asymptotically free.
Too large coupling constants result in the Landau-pole problem below the GUT scale.
In order to avoid the Landau-pole problem, 
we put $k^{(d)}=k^{(\ell)}=4\pi$ at the GUT scale as the upper bound on the coupling constants.%
\footnote{
The perturbative analysis is no more viable for $k={\cal O}(4\pi)$.
Our result, however, does not strongly depend on the values of the coupling 
constants at the GUT scale as long as they are large.
}
With this boundary condition, we obtain $k^{(d)}=0.99$,~$k^{(\ell)}=0.74$ and $m^{(d)}/m^{(\ell)} =0.71$ at the mediation scale in the minimal R-invariant gauge mediation model for $m_{3/2}={\cal O}(1)-{\cal O}(10)$\,keV.%
\footnote{
As we see shortly, the mediation scale is required to be close to $\sqrt{F}$
to obtain heavy enough gaugino masses, which is determined for a given gravitino mass.
}
The explicit form of the RG equations are given in Ref.~\cite{Sato:2009dk}.
In this analysis, we have neglected the contribution of the SUSY breaking sector to the RG equations, which may make the values of $k^{(\chi)}$ a little smaller.


In Fig.\,\ref{fig:1kev_z}, we show the SSM mass spectrum as  a function of the parameter $\langle S\rangle /m^{(\ell)}$ for 
 $m_{3/2} = 1~\KEV$ and $m_{3/2} = 10~\KEV$.
We fixed $k^{(\ell)} F/m^{(\ell)2} = 0.9$ in both figures.
The figures show that the sfermion masses are  much heavier than the gauginos.
The sfermion-gaugino mass ratio is larger for the heavier gravitino mass if we fix the order of the gaugino masses.%
\footnote{
In Refs.\cite{Ibe:2005xc,Sato:2010tz},
the SSM spectrum in the R-invariant gauge mediation with the very light gravitino with a mass 
below 16\,eV has been considered.
There, the sfermion-gaugino mass ratio is much smaller.
}
The figure also shows that the gaugino masses are maximized when $k^{(\ell)} \langle S \rangle \sim m^{(\ell)}$.
In Fig. \ref{fig:1kev}, we also show the mass spectrum of gaugino and sfermion as a function of the parameter $k^{(\ell)} F/m^{(\ell)2}$.
In the figure, we took $\langle S\rangle$ as a value which maximizes the gluino mass.

\begin{figure}[tbp]
\begin{center}
\begin{minipage}{.46\linewidth}
\includegraphics[width=\linewidth]{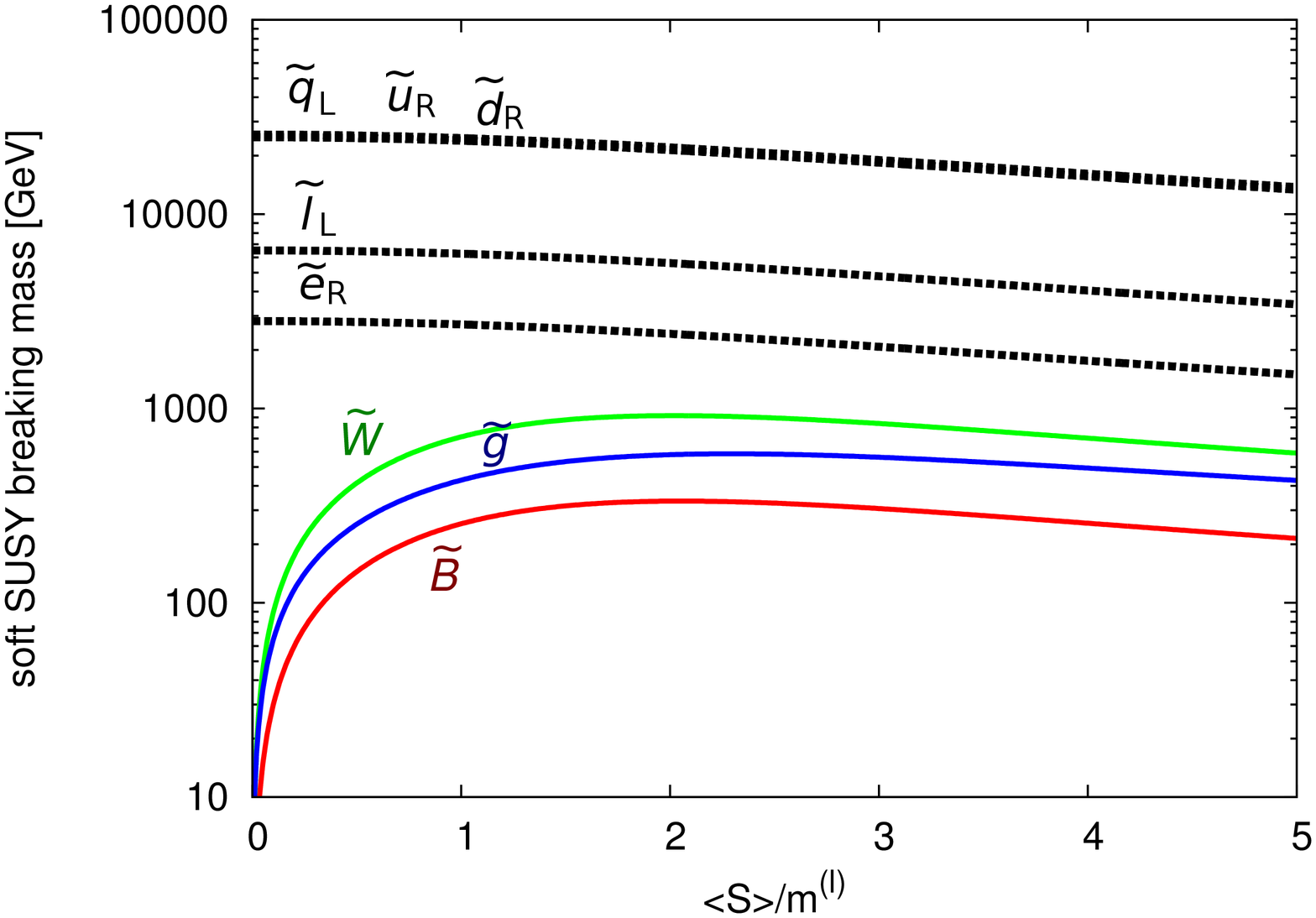}
\end{minipage}
\hspace{.02\linewidth}
\begin{minipage}{.46\linewidth}
\includegraphics[width=\linewidth]{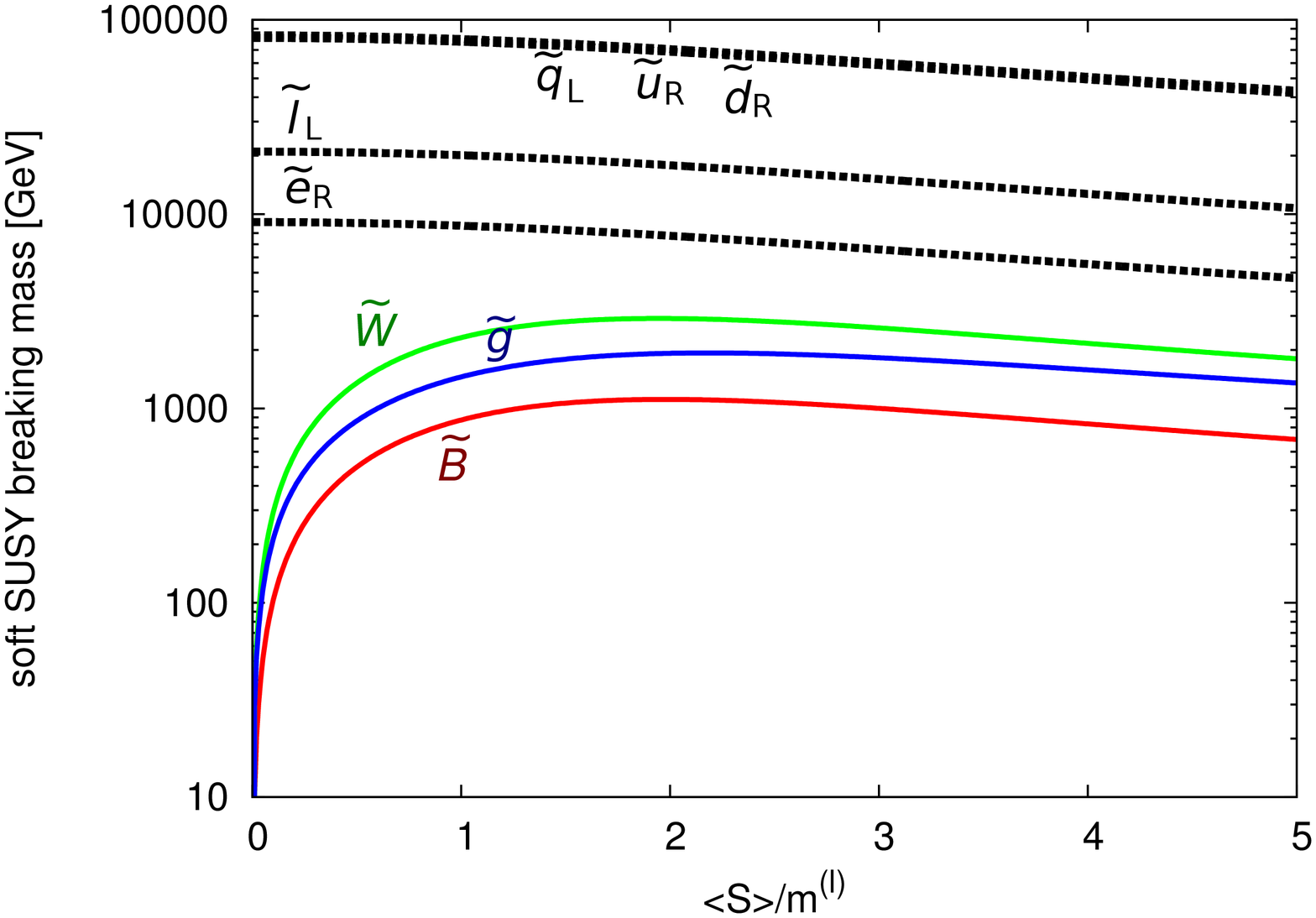}
\end{minipage}
\caption{
The soft SUSY breaking masses of gauginos and sfermions in the R-invariant model with double messenger
for $m_{3/2}=1$\,keV (left) and $m_{3/2}=10$\,keV (right).
We set $k^{(\ell)} F/m^{(\ell)2}=0.9$.
}\label{fig:two_1kev_z}
\vspace{2cm}
\begin{minipage}{.46\linewidth}
\includegraphics[width=\linewidth]{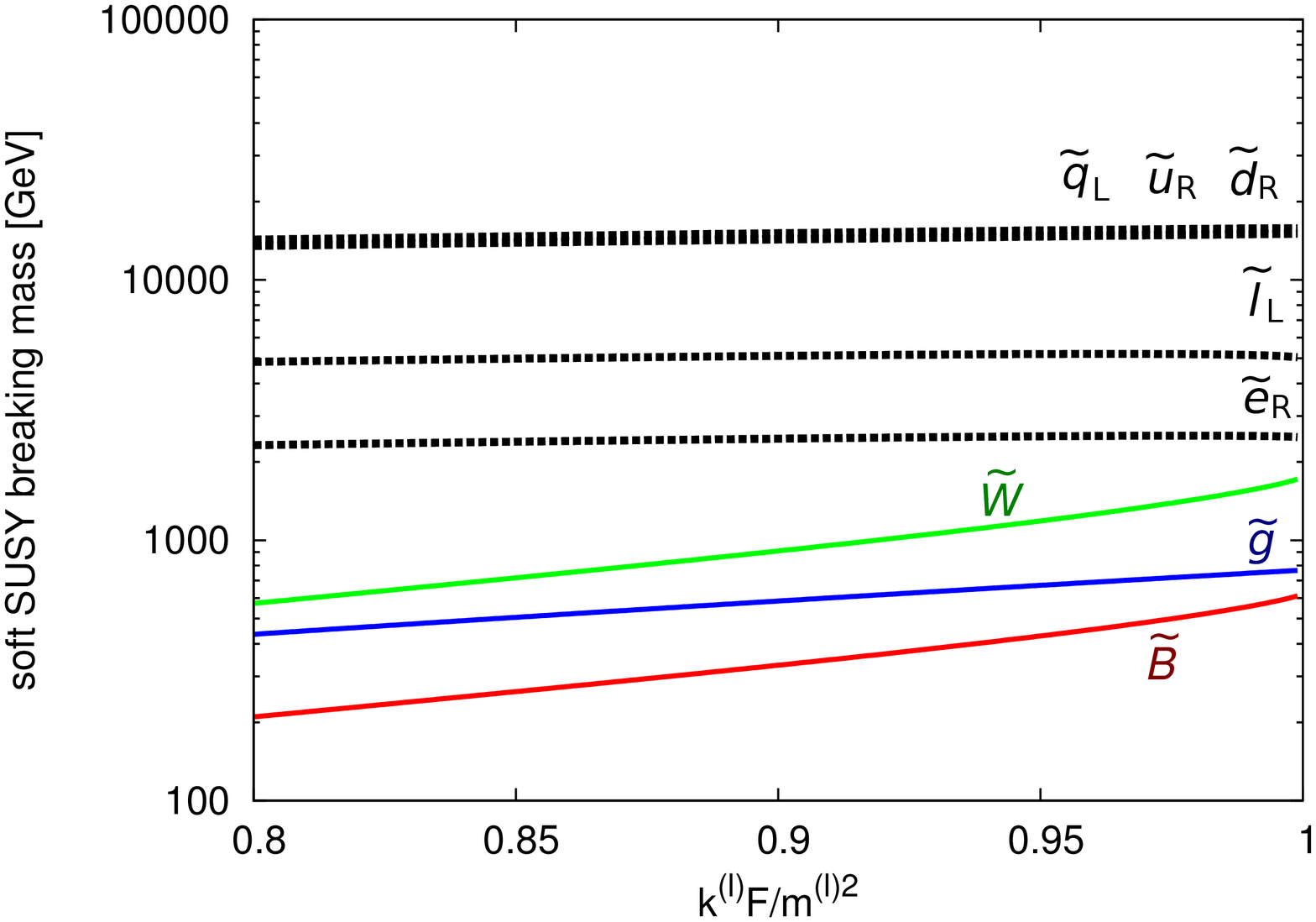}
\end{minipage}
\hspace{.02\linewidth}
\begin{minipage}{.46\linewidth}
\includegraphics[width=\linewidth]{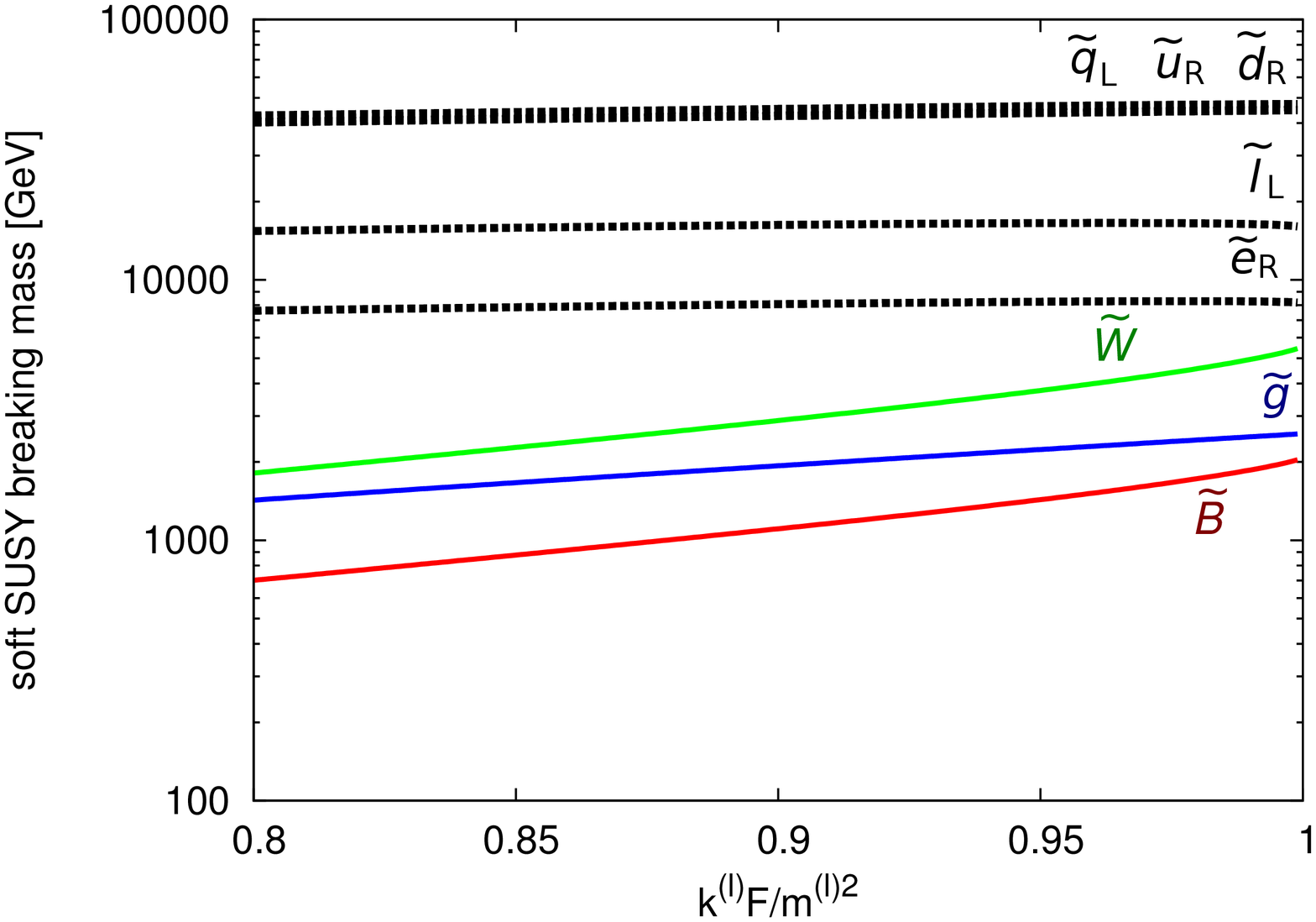}
\end{minipage}
\caption{
The soft SUSY breaking masses of gaugino and sfermion in the R-invariant model with double messenger
for $m_{3/2}=1$\,keV (left) and $m_{3/2}=10$\,keV (right).
We take $\langle S\rangle$ as a value which maximizes the gluino mass.
}\label{fig:two_1kev}
\end{center}
\end{figure}

The interesting observation here is that the gluino mass is predicted to be rather light 
and can be lighter than the wino even though we have assumed the GUT boundary conditions.
Thus, the gaugino mass spectrum is distinguishable from the one in the minimal gauge mediation,
$M_{\rm bino} : M_{\rm wino} : M_{\rm gluino}\simeq 1:2:6$.

In Fig.\,\ref{fig:two_1kev_z} and \ref{fig:two_1kev}, we also show the SSM mass spectrum for the model
with two additional pairs of massive messengers $\Psi_{i}^\prime$, $\tilde\Psi_i^\prime$ $(i=1,2)$
which have the masses and the couplings to $S$ similar to those of $\Psi_{i}$ and $\tilde\Psi_i$
in Eq.\,(\ref{eq:messenger}).
In the next section, we consider the SUSY breaking model where the R-symmetry is 
spontaneously broken in a perturbative way with the help of the $U(1)$ gauge interaction.
In that model, the messenger sector is required to be doubled (see also Eq.\,(\ref{eq:messenger2})).
The sfermion-gaugino mass ratio is smaller than the minimal model,
because the gaugino mass is proportional to the messenger flavor number
but the sfermion mass is proportional to square root of the messenger flavor number.

The figure shows that the gluino mass is more close to the bino mass in the doubled messenger model.
For such a peculiar spectrum, the collider phenomenology can be significantly 
different from the usual models with the gaugino masses which satisfies GUT relations,
and hence, may require different search strategies than the usual SUSY scenarios 
(see for example Ref.\,\cite{Ajaib:2010ne}).
The detailed collider study for the above gaugino spectrum will be given elsewhere\,\cite{Future}.

\section{Entropy production from SUSY breaking sector}\label{sec:Entropy}
In the previous section, we have discussed the mass spectrum in the minimal R-invariant gauge mediation model, and the peculiar mass spectrum are predicted for the gravitino mass 
in the one to ten keV range.
As we have mentioned in the introduction, however, the relic density of the  gravitino with a mass
in this range is too high to be consistent with the observed dark matter density
if the gravitinos were in the thermal equilibrium in the early universe.
The thermally produced gravitino density is roughly given by,
\begin{eqnarray}
 \Omega_{3/2} h^2 \simeq 0.1\times \left(\frac{100}{g_*(T_D)}\right)
 \left(\frac{m_{3/2}}{100\,{\rm eV}}\right)\ ,
\end{eqnarray}
where $g_*(T_D)\simeq 100$ denotes the effective massless degree of freedom 
in the thermal bath at the decoupling temperature, $T_D$, of the gravitino
from the thermal bath\,\cite{Moroi:1993mb}, 
\begin{eqnarray}
\label{eq:decouple}
 T_D \sim \max\left[\, M_{\rm gluino}, \,\,
 26\,{\rm GeV}\left(
 \frac{g_*(T_D)}{100}
 \right)^{1/2}
 \left(\frac{m_{3/2}}{1\,{\rm keV}}\right)^2
 \left(
 \frac{500\,{\rm GeV}}{M_{\rm gluino}}
 \right)^2
 \right]\ .
\end{eqnarray}
Notice that the effective interactions between the gravitino and the SM fermions 
after integrating the SUSY particles out are so suppressed that 
they cannot keep the gravitino in the thermal bath after the gauginos decouple.
In the above expressions, we have neglected the contributions from
the Winos and Binos which could give comparable or even a larger contributions.
The following discussion is  not affected as long as the order of the magnitude 
of $T_D$ is not changed.%
\footnote{
One may obtain more precise expressions of $T_D$ by using
more recent analysis on the gravitino production cross section given in 
Ref.\,\cite{Bolz:2000fu}.
}

From the above discussion, 
for the gravitino with a mass in the one to ten keV range to be a consistent dark matter candidate,
the above relic density should be diluted by 
\begin{eqnarray}
\Delta \simeq 100\times \left(\frac{100}{g_*(T_D)}\right)\left(\frac{m_{3/2}}{10\,\rm keV}\right)\ ,
\end{eqnarray}
after the decoupling of the gravitino.%
\footnote{
The gauge mediation mechanism often involves natural mechanisms of 
late time entropy production from, for example,  the messenger sector\,\cite{Fujii:2002fv}
or the intermediate SUSY breaking sector\,\cite{Fujii:2003iw}. 
}
It should be noted that the late time entropy production dilutes also the primordial baryon asymmetry by the same factor, but it may not cause any serious problem\,\cite{Fujii:2002fv} in the thermal 
leptogenesis\,\cite{Fukugita:1986hr} for $\Delta \simeq 10-100$.
 
In the previous section,  we have also made a tacit but a crucial assumption; 
the spontaneous R-symmetry breaking, $\vev S\neq 0$.
It is not trivially realized in many SUSY breaking models. 

In this section, we propose a very ambitious solution to both the above problems,
the dilution of the thermal gravitino and the R-symmetry breaking,
at the same time, by considering the entropy production from the SUSY breaking sector,
where the spontaneous R-symmetry breaking is realized.

\subsection{Extended vector-like SUSY breaking sector}
As an example of the SUSY breaking model where
the spontaneous R-symmetry breaking is realized,
we consider the extended model of the vector-like dynamical SUSY breaking model based on 
$SU(2)$ gauge theory in Ref.\,\cite{Izawa:1996pk,Intriligator:1996pu}.
In the extended model, one of the global $U(1)$ symmetry 
is upgraded to a gauge symmetry\,\cite{Dine:2006xt}
which is spontaneously broken at the SUSY breaking vacuum.
As discussed in Ref.\,\cite{Ibe:2009dx}, the spontaneous R-symmetry breaking 
is achieved in a perturbative way, which 
is not the case in the original SUSY breaking model.

The notable property of the extended model is 
that it possesses an accidental discrete symmetry even after the spontaneous 
$U(1)$ symmetry breaking\,\cite{Ibe:2009dx}. 
Thus, the lightest particle which is charged under the unbroken discrete symmetry 
has a long lifetime.
As we will show shortly, the energy density of such a long lived particle can 
dominate the universe and cause the entropy production when it decays,
which dilutes the thermally produced  gravitino.

The vector-like SUSY breaking model consists of four fundamental representations
$Q_k\, (k=1,\cdots,4)$ and six singlets $S_{ij}=-S_{ji}\, (i,j=1,\cdots,4)$ which interact 
with each other in the superpotential,
\begin{eqnarray}
 W = \sum\lambda_{ij}^{kl} S_{ij}Q_kQ_l\ ,
\end{eqnarray}
where $\lambda$'s denote the coupling constants. 

In the extended model, we gauge one of the $U(1)$ subgroup of the maximal 
subgroup of $SU(4)$ of the model
 by assigning
the gauge charges, $Q_{1,2}(1/2)$, $Q_{3,4}(-1/2)$, $S_{12}(-1)$, $S_{34}(+1)$, and $S_{13,14,23,24}(0)$.
With this charge assignment, the above superpotential reduces to  
\begin{eqnarray}
\label{eq:tree}
W = \lambda^{(+)} S_{12} Q_1 Q_2 + \lambda^{(-)} S_{34}Q_3 Q_4 + 
\sum {\lambda^\prime}_{ij}^{kl} S_{ij} Q_kQ_l\ ,
\end{eqnarray}
where ${\lambda^\prime}_{ij}^{kl} = 0 $ for $ij = 12, 34$ or $kl=12,34$.
The global symmetries of the model are $U(1)_R\times Z_4$. 
The charge assignment of the R-symmetry is $S(2)$ and $Q(0)$.
The fields are transformed to $i Q $ and $-S$ under the $Z_4$ symmetry.%
\footnote{Classically, the $Z_4$ symmetry can be realized as a continuous $U(1)$
symmetry with the charge assignment $S(2)$ and $Q(-1)$.
The anomaly against the $SU(2)$ gauge symmetry breaks the $U(1)$ symmetry
down to the discrete $Z_4$ subgroup.}
We list the symmetries of the model in Table\,\ref{tab:symmetries}.
\begin{table}[t]
\caption{The symmetries of the model. $SU(2)\times U(1)$ 
are gauge symmetries and $Z_4\times U(1)_R$ are global symmetries.
Notice that both the global symmetries are anomaly free.}
\begin{center}
\begin{tabular}{|c|c|c||c|c|}
\hline
& $SU(2)$ & $U(1)$ & $Z_4$& $U(1)_R$\\
\hline
$S_{12}$& $\bf 1$ & $-1$ & $e^{i\pi }$ & $2$\\
\hline
$S_{34}$& $\bf 1$ & $1$ & $e^{i\pi }$ & $2$\\
\hline
$S_{13,14,23,24}$& $\bf 1$ & $0$ & $e^{i\pi}$ & $2$\\
\hline
$Q_{1,2}$& $\bf 2$ & $1/2$ & $e^{i\pi/2 }$ & $0$\\
\hline
$Q_{3,4}$& $\bf 2$ & $-1/2$ & $e^{i\pi/2 }$ & $0$\\
\hline
\end{tabular}
\end{center}
\label{tab:symmetries}
\end{table}%

Below the dynamical scale $\L$, the model is well described by using 
the composite fields, $M_{ij} \sim Q_{i}Q_{j}/\L$, 
whose superpotential terms are approximated by,
\begin{eqnarray}
\label{eq:eff}
 W_{\rm eff} = \lambda^{(+)} \Lambda S_{12} M_{12} + \lambda^{(-)}\L S_{34} M_{34}
+ \sum {\l^\prime}^{kl}_{ij} \L S_{ij} M_{kl}\ 
 +{\cal X}({\rm Pf}(M_{ij}) - \Lambda^2)\ ,
\end{eqnarray}
where $\cal X$ denotes the Lagrange multiplier which expresses the quantum deformed 
moduli constraint, ${\rm Pf}(M)=\L^2$.
Here, the ambiguity of the normalizations of the meson fields are implicitly absorbed by $\l$'s.
Thus, strictly speaking, the parameters $\l$'s appearing in Eqs.\,(\ref{eq:eff}) and (\ref{eq:eff2})
are different from the ones in Eq.\,(\ref{eq:tree}).
By using appropriate linear combinations of $S$'s and $M$'s we may rewrite the above 
effective field theory by,
\begin{eqnarray}
\label{eq:eff2}
W = \lambda^{(+)} \L S_+ M_- + \lambda^{(-)} \L S_-M_+  +\sum_{a = 1\cdots4} \l^{\prime}_a \L S_a M_a
+ {\cal X}\left( M_+ M_-  +\sum_{a,b=1\cdots4}\frac{ y_{ab}}{2} M_a M_b - \L^2 \right).
\end{eqnarray}
Here, the newly introduced 
matrix $y_{ab}$ is generically given by,
\begin{eqnarray}
 y_{ab} = (U^{T} U)_{ab}, \quad U\in SU(4)\ .
\end{eqnarray}

By assuming that $\lambda$'s are perturbative,
and $\l^{\pm}$ are smaller than $\l^\prime$'s,
we may parametrize the deformed moduli space,
by,%
\footnote{The ``radial" component of the $M_{\pm}$ becomes
a ``mass partner'' of ${\cal X}$, and hence, we can integrate the radial component out. 
}
\begin{eqnarray}
 M_+ = e^{\phi/\sqrt{2} \L} \sqrt{\L^2 -  \sum_{a,b=1\cdots4}\frac{ y_{ab}}{2} M_a M_b} 
 \ , 
  \quad  M_- = e^{-\phi/\sqrt{2} \L}\sqrt{\L^2 -  \sum_{a,b=1\cdots4}\frac{ y_{ab}}{2} M_a M_b} 
  \ . 
\end{eqnarray}
Notice that the $U(1)$ gauge symmetry is spontaneously broken on the deformed moduli space.
Then, the above effective theory can be reduced to
\begin{eqnarray}
W =  \left(\l^{(+)}\L S_+ e^{-\phi/\sqrt{2}\L} + \l^{(-)}\L S_- e^{\phi/\sqrt{2}\L} \right) 
 \sqrt{\L^2 -  \sum_{a,b=1\cdots4}\frac{ y_{ab}}{2} M_a M_b}
 +\sum_{a = 1\cdots4} \l^{\prime}_a \L S_a M_a\ .
\end{eqnarray}
In the followings, we simplify the model by taking $\l^{(\pm)}=\l$ and $y_{ab} = \d_{ab}$,
although we can easily generalize the results for more generic coupling constants.%

By expanding the above superpotential around $\phi=0$ and $M_{a} = 0$, we obtain,
\begin{eqnarray}
\label{eq:superST}
 W &=& \l \L^2(S_+ + S_-) - \frac{\l}{\sqrt{2}} \L (S_+ - S_-) \phi
 + \frac{\l}{4}(S_++S_-) \left(\phi^2 - \sum_{a =1\cdots 4}M_a^2\right) 
 +\sum_{a = 1\cdots4} \l^{\prime}_a \L S_a M_a\ , \cr
 &=& \sqrt 2\l \L^2S - \l \L T \phi
 + \frac{\l}{2\sqrt 2}S\left(\phi^2 - \sum_{a =1\cdots 4}M_a^2\right) 
 +\sum_{a = 1\cdots4} \l^{\prime}_a \L S_a M_a\ , 
\end{eqnarray}
where we have defined $S$ and $T$ by,
\begin{eqnarray}
 S  =  \frac{1}{\sqrt 2}(S_+ + S_-)\ , \quad
 T =  \frac{1}{\sqrt 2}(S_+ - S_-) \ .  
 \end{eqnarray}
The newly defined $S$ corresponds to the pseudo-flat direction
which $F$-term breaks SUSY. 
 
The above effective theory 
possesses an discrete symmetry, $Z_2$,
after the SUSY and the $U(1)$ gauge symmetry breaking,
which is a diagonal subgroup of the discrete subgroup of the $U(1)$ gauge symmetry
and the global $Z_4$ symmetry.
Under the $Z_2$ symmetry, $M_a$ and $S_a$ are odd while the other $S$, $T$, and $\phi$ are even
(see Table\,\ref{tab:symmetries2}).%
\footnote{
In addition to the $Z_2$ symmetry, the model apparently possesses 
another discrete symmetry under which $\phi$ and $T$ are odd.
This symmetry stems from the charge conjugation symmetry
of the gauged $U(1)$ symmetry.
The charge conjugation symmetry is, however, expected 
to be broken in generic models with $\l^{(+)}\neq \l^{(-)}$.
In the followings, we do not consider the charge conjugation
symmetry as a good symmetry.
}
Therefore, the lightest particle which is odd under $Z_2$ symmetry is stable.
In the appendix, we give the mass spectrum of the SUSY breaking sector which shows
that the lightest $Z_2$ odd particle is an appropriate linear combination of the scalar components of
$M_a$ and $S_a$.
In the followings, we name the lightest $Z_2$ odd particle, $x_L$.

\begin{table}[t`]
\caption{The symmetries of the model in terms of the low energy fields. 
By the condensation, $\vev{M_+M_-}=\L^2$, 
the symmetries $U(1)\times Z_4$ break down to a
global $Z_2$ symmetry under which $S_a$ and $M_a$ are odd.
}
\begin{center}
\begin{tabular}{|c|c||c|c|}
\hline
& $U(1)$ & $Z_4$& $U(1)_R$\\
\hline
$S_{-}$& $-1$ & $e^{i\pi }$ & $2$\\
\hline
$S_{+}$& $1$ & $e^{i\pi }$ & $2$\\
\hline
$S_{a}$& $0$ & $e^{i\pi}$ & $2$\\
\hline
$M_{-}$& $-1$ & $e^{i\pi }$ & $0$\\
\hline
$M_{+}$& $1$ & $e^{i\pi }$ & $0$\\
\hline
$M_{a}$& $0$ & $e^{i\pi}$ & $0$\\
\hline

\end{tabular}
\end{center}
\label{tab:symmetries2}
\end{table}%

\subsection{Decay of the $Z_2$ charged particle}
So far, we have assumed that the $Z_2$ accidental symmetry  is an exact symmetry.
Generically, such an accidental symmetry could be broken by higher dimensional operators 
which are suppressed by the reduced Planck scale.
For example, the lowest dimensional operator which breaks the $Z_2$ symmetry is
\begin{eqnarray}
\label{eq:Z2b}
 W \sim \frac{c}{M_{P}} QQ H_u H_d \sim \frac{c \L}{M_{P}}M H_u H_d \ ,
\end{eqnarray}
where $c$ is a coupling constant and $H_u$ and $H_d$ denote the Higgs doublets in the SSM.
Here, we have assumed that the $R$-charge of $H_u H_d$ is 2, assuming that the R-symmetry
(or  its discrete subgroup) is better symmetry than the accidental $Z_2$ symmetry.

Through this operator, the lightest $Z_2$ odd particle 
decays into Higgs (Higgsino) with the decay rate
\begin{eqnarray}
\G_{x_L} \sim  \frac{1}{8\pi} \frac{c^2 \L^2}{M_{P}^2}m_{x_L}\ .
\end{eqnarray}
Here, we have neglected the masses of the Higgs and Higgsinos in the final state.
As a result, the decay temperature is roughly given by,%
\footnote{The detailed analysis is given in the appendix.}
\begin{eqnarray}
\label{eq:Tdecay}
 T_{\rm decay} \sim \left(\frac{90}{\pi^2 g_*}\right)^{1/4} \sqrt{ \G_{x_L} M_{P}}
 \sim 1\,{\rm GeV}\times c\left(\frac{10}{g_*}\right)^{1/4}\left(\frac{\L}{10^7\,\rm GeV}\right)
 \left(\frac{m_{x_L}}{10^6\,\rm GeV}\right)^{1/2}\ .
\end{eqnarray}
Therefore, the decay temperature of the lightest $Z_2$ odd particle is expected to be very low
even if the $Z_2$ symmetry is explicitly broken by the higher dimensional operators.

\subsection{The entropy production and dilution of the gravitino}
\begin{figure}[t]
\begin{center}
 \begin{minipage}{.45\linewidth}
  \includegraphics[width=\linewidth]{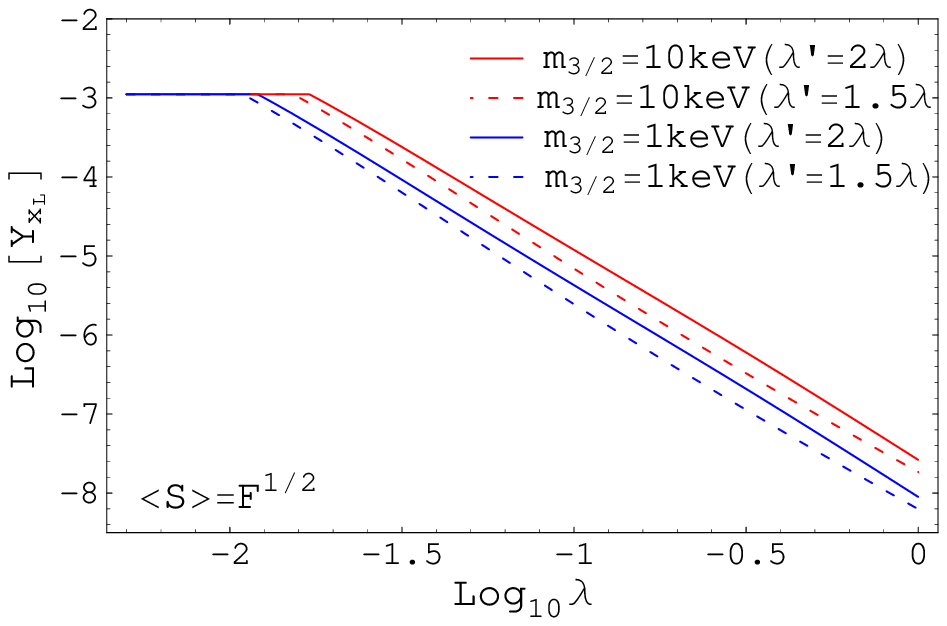}
 \end{minipage}
 \begin{minipage}{.43\linewidth}
  \includegraphics[width=1.0\linewidth]{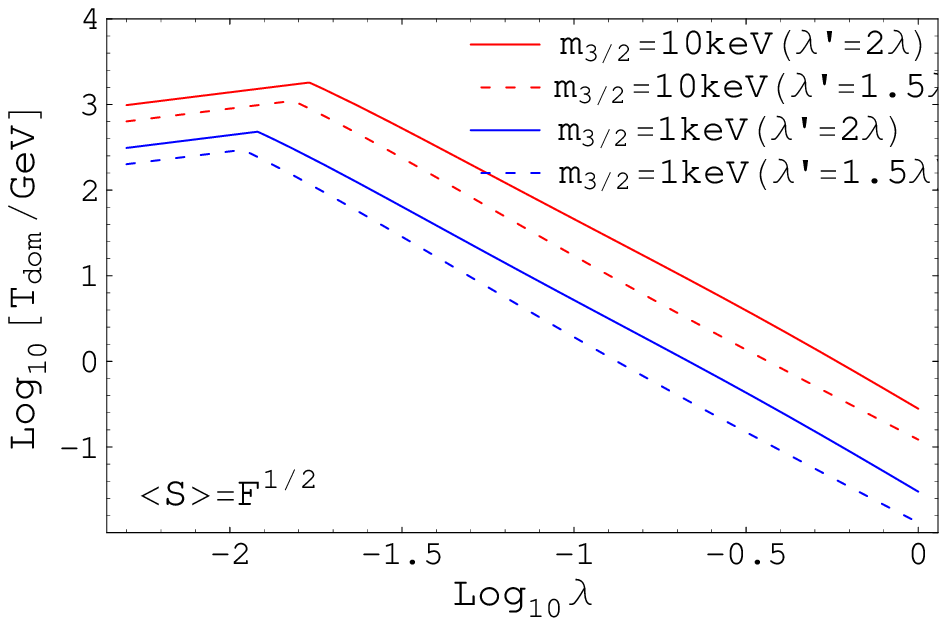}
 \end{minipage}
 \end{center}
\caption{
Left) The yield of the lightest $Z_2$ odd particle, $x_L$, for a given SUSY breaking scale.
Right) The domination temperature of the lightest $Z_2$ charged particle after freeze out.
In the figure, we have used $g_*\simeq 250$ at the time of the freeze-out of $x_L$.}
\label{fig:Tdom}
\end{figure}
Since the decay temperature of the lightest $Z_2$ odd particle is very low,
the energy density of the lightest $Z_2$ odd particle is 
expected to dominate the universe, where the domination temperature is estimated by,
\begin{eqnarray}
 T_{\rm dom}\simeq \frac{4}{3}m_{x_L} Y_{x_L}\ ,\label{eq:domtemperature}
\end{eqnarray}
where $Y_{x_L}$ is the yield (number density divided by entropy density) of $x_L$ after its freeze-out.
The yield is roughly given by,
\begin{eqnarray}
  Y_{x_L} \simeq \min\left[ \frac{0.278}{g_*},  \frac{0.76}{g_*^{1/2} M_{P}T_f\vev{ \sigma v_{\rm rel} }}
  \right]\ ,
\end{eqnarray}
where $g_*$ is the effective massless degree of freedom at the freeze-out temperature $T_f$, and the $\vev{\sigma v_{\rm rel}}$ is the thermal averaged annihilation cross section,
\begin{eqnarray}
\vev{\sigma v_{\rm rel}} \sim \frac{1}{8\pi}\frac{\l^4}{m_{x_L}^2}\ .
\end{eqnarray}
As we show in the appendix, 
the dominant mode of the annihilation process of $x_L$
is the one into a pair of the gravitinos.%
\footnote{The gravitinos produced at the annihilation of $x_L$ 
interact with the thermal bath and are thermalized 
immediately since $T_f \gg T_{D}$.}

\begin{figure}[t]
\begin{center}
  \includegraphics[width=.5\linewidth]{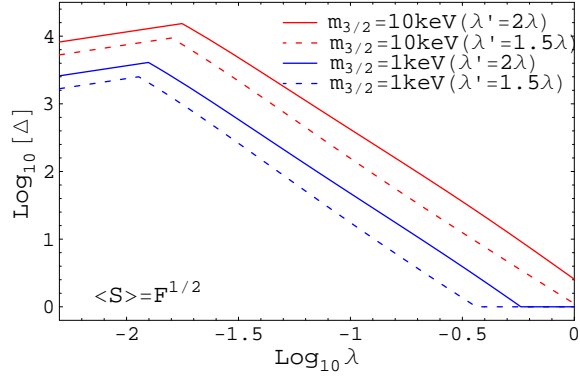}
  \end{center}
\caption{
The dilution factor for a given SUSY breaking scale.
In the figure, we have  fixed $T_{\rm decay}\simeq 100$\,MeV 
by choosing an appropriate coefficient $c$ in Eq.\,(\ref{eq:Tdecay}).}
\label{fig:Entropy}
\end{figure}

In Fig.\,\ref{fig:Tdom}, we show the yield and the domination temperature.
The figure shows that the domination temperature 
is typically below the decoupling temperature of the gravitino given in Eq.\,(\ref{eq:decouple}).
Therefore, after the decay of $x_L$, the yield of the gravitino is diluted by the factor $\D^{-1}$, 
\begin{eqnarray}
\D &\equiv& \frac{s|_{\rm after~decay}}{s|_{\rm before~decay}}\simeq \frac{4}{3T_{\rm decay}}\frac{\rho|_{\rm after~decay}}{s|_{\rm before~decay}}
\sim \frac{4}{3T_{\rm decay}}\frac{m_{x_L}n_{x_L}|_{\rm before~decay}}{s|_{\rm before~decay}}
\simeq \frac{4}{3T_{\rm decay}}m_{x_L}Y_{x_L}\, \nonumber 
\end{eqnarray}
where $s$ is the entropy density, $\rho$ is the radiation energy density, and $n_{x_L}$ is the number density of $x_L$. 
Thus we obtain (using Eq.~(\ref{eq:domtemperature}))
\begin{eqnarray}
\D \sim  \frac{T_{\rm dom}}{T_{\rm decay}}\ .
\end{eqnarray}
In Fig.\,\ref{fig:Entropy}, we show the resultant
dilution factor for a given SUSY breaking scale.
In the figure,
we assumed $T_{\rm decay}\simeq 100$\,MeV
which roughly corresponds to the lower bound on the decay temperature 
not to affect the Big-Bang Nucleosynthesis.%
\footnote{The decay temperature, $T_{\rm decay}\simeq 100$\,MeV, can be achieved by 
choosing appropriate coefficient $c$.}
The figure shows that the dilution factor of the order of $10-100$ 
is achieved  for $\l \sim O(10^{-1})$, 
which is required to achieve the consistent gravitino dark matter scenario
with a mass in the one to ten keV range.

\subsection{Gravitino from the process of entropy production}
The thermal relic gravitinos are diluted as we have discussed above.
In this subsection, we discuss other sources of gravitinos. 
The decay of $x_L$ produces higgsinos with a branching ratio of order 1, and they eventually decay into gravitinos after some cascade decay.
Besides, the particles from the $x_L$ decay have very large energies of order $m_{x_L}$, and they interact with the 
thermal bath and may produce more SUSY particles.

Suppose that $N_{x_L}$ gravitinos are produced in average from a decay of single $x_L$. Then the yield of the gravitinos coming from the decay is,
\beq
Y_{3/2}^{\rm decay} \sim N_{x_L} \frac{n_{x_L}}{s|_{\rm after~decay}} \sim \frac{N_{x_L}}{m_{x_L}}\frac{\rho|_{\rm after~decay}}{s|_{\rm after~decay}}
\sim \frac{3}{4}N_{x_L} \frac{T_{\rm decay}}{m_{x_L}}.
\eeq
Thus, the contribution of these gravitinos to $\Omega h^2$ is,
\beq
\Omega^{\rm decay}_{3/2} h^2 
\sim 2\times 10^{-4} \times N_{x_L}  \left(\frac{m_{3/2}}{10~\KEV}\right) \left(\frac{T_{\rm decay}}{100~\MEV}\right) \left(\frac{10^6~\GEV}{m_{x_L}}\right).
\eeq

The decay remnants of $x_L$ whose initial energies are of the order of $E\simeq m_{x_L}$
lose most of their energies by interacting with the particles in the thermal bath.
Then, the SUSY particle productions take place when 
the energies of the remnants decrease down to the threshold energy, 
$E\simeq M_{\rm gaugino}^2/T_{\rm decay}$.
As a result, the average number of the gravitinos from the decay of $x_L$ is 
at most of the order of unity\,\cite{Fujii:2002fv}.

Here, instead of analyzing the detail of the above process,
we give a very rough upper bound on $N_{x_L}$ by neglecting 
the energy loss of the remnants, which is good enough for our discussion. 
That is, the maximum number of the SUSY particles from a decay of $x_L$
is achieved if the initial energy $E\simeq m_{x_L}$ is distributed to 
$\bar{N}\simeq m_{x_L}/(M_{\rm gaugino}^2/T_{\rm decay})$ particles with energy 
$E\simeq M_{\rm gaugino}^2/T_{\rm decay}$, and each particle with this energy produces a gaugino.
Thus, the absolute upper bound on the number $N_{x_L}$ is given by,
\beq
N_{x_L} \lsim\bar{N}&\simeq& 10\cdot\left(\frac{100~\GEV}{M_{\rm gaugino}}\right)^2\left(\frac{T_{\rm decay}}{100~\MEV}\right)\left(\frac{m_{x_L}}{10^6~\GEV}\right)+2{\rm Br}(x_L \to \tilde{H}_u \tilde{H}_d), 
\eeq
where the second term comes from the direct higgsino production by the decay of $x_L$.
Therefore, we find that the gravitino coming from 
the $x_L$ decay does not become a dominant component of the dark matter.%
\footnote{Although the energy density of the gravitino component from the decay of $x_L$ is subdominant, such gravitinos may have much larger velocity than the one of the thermally 
produced gravitino.
Thus, the gravitino component from the decay of $x_L$ 
could have some impacts on the structure formation.
}

\subsection{Spontaneous R-symmetry breaking}
Finally, we discuss the R-symmetry 
breaking in the extended vector-like SUSY breaking model.
As we see from the superpotential in Eq.\,(\ref{eq:superST}),
the SUSY breaking field $S$ has the flat potential at the tree level,
and hence, $S$ corresponds to the pseudo-flat direction.
The potential of $S$ is, however, deformed by the radiative correction,
and especially, the origin of $S$ can be destabilized\,\cite{Dine:2006xt}. 

At the one-loop level, the effective potential of the pseudo-flat direction is given by
the so-called Coleman-Weinberg potential,
\begin{eqnarray}
\label{eq:CW}
 V_{\rm CW}(S) = \frac{1}{64\pi^2}\tr (-)^FM^4(S) \log \frac{M^2(S)}{\m_R^2}\ ,
\end{eqnarray}
where $\m_R$ denotes the renormalization scale, and $(-)^F = 1$ for bosons and $(-)^F = -1$ for fermions.
The mass spectrum in the extended model 
is given in the appendix.

In Fig.\,\ref{fig:Rbreaking}, we show the parameter space where 
the spontaneous R-symmetry breaking is achieved for given values of $\lambda$.
In the figure, $g_X$ denotes  the gauge coupling constant of the $U(1)$ gauge interaction.
The figure shows that  spontaneous R-symmetry breaking is realized for
$g_{X}\gsim \l, k^{(\ell)}$, 
which corresponds to the large contribution
from the $U(1)$ gauge interaction.
Notice that for $g_X\gsim 0.5$ at the mediation scale, the $U(1)$ gauge 
interaction has the Landau pole below the GUT scale.
Thus, if we require that the extended model is perturbative up to the GUT scale,
the gauge coupling constant should satisfy $g_X\lsim 0.5$ at the mediation scale.

For the larger field value of $S\gg M_{\rm med}$, 
the radiatively generated potential is further approximated by~\cite{ArkaniHamed:1997ut},
\begin{eqnarray}
 V_{\rm CW}(S) &=& \frac{|F|^2}{Z_{S}}\ , \cr
  Z(S) &=&   \exp\left[-\int_{\ln M_{\rm med}}^{\ln S}2\gamma_S\, d\ln\mu\right]\ .
\end{eqnarray}
Here, $\g_S$ is the anomalous dimension of $S$ which is, at the one-loop level, given by,
\begin{eqnarray}
 2\g_S = \frac{4}{2\pi} \frac{k^{(\ell)2}}{4\pi}+ \frac{6}{2\pi} \frac{k^{(d)2}}{4\pi}
-  \frac{1}{\pi} \frac{g_X^2}{4\pi}+{\cal O}(\lambda^2)\ ,
\end{eqnarray}
where ${\cal O}(\lambda^2)$ contribution comes from mesons $M$ in the region $\l S \lsim \L$ or quarks $Q$ in the region $\l S \gsim \L$, and
this contribution is always positive in the region where perturbative calculation is valid (i.e. $\l S$ is not near $\L$).
In order for the potential not to show the runaway behavior, 
we need to have $\g_S > 0$ at the larger value of $S$.
Because ${\cal O}(\l^2)$ contribution is positive, we can conservatively neglect this contribution to constrain the viable parameter space. 
So we neglect this contribution in $\g_S$.
In Fig.\,\ref{fig:Rbreaking}, we have shown the region where $\gamma_S > 0$ at least
around the mediation scale.
The figure shows that spontaneous R-breaking is realized for $g_{X}\gsim \l, k^{(\ell)}$,
while the runaway behavior is avoided for not too large $g_X$ compared with $k^{(\ell)}$.%
\footnote{
In the figure, we have fixed $\sqrt{k^{(\ell)}F}/m^{(\ell)} = 0.8$.
For the heavier messenger masses, the hatched regions in the figure are shifted 
to right while the light-shaded region is not shifted. 
This is because the contribution to the Coleman-Weinberg potential in Eq.\,(\ref{eq:CW})
from the messenger is suppressed by the heavier masses.
Thus, for the heavier messengers, the allowed parameter space is larger.
}

\begin{figure}[t]
\begin{center}
  \includegraphics[width=.5\linewidth]{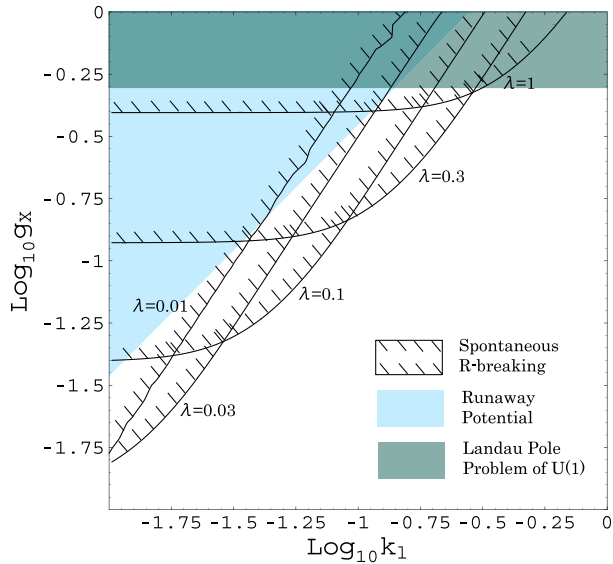}
  \end{center}
\caption{
Spontaneous R-symmetry breaking is realized by the radiatively generated 
potential of $S$ in the hatched region for given values of $\lambda$.
The $U(1)$ gauge coupling $g_X$ and the messenger-SUSY breaking field
coupling $k^{(\ell)}$ are defined at the mediation scale.
In the figure, we have fixed $\l^\prime =2\l$, $k^{(d)} \simeq 1.8 k^{(\ell)}$,
$m^{(d)}=2.1m^{(\ell)}$, and $m^{(\ell)}=\sqrt{k^{(\ell)}F}/0.8$.
We have fixed the value of $F$.
In the dark shaded region, the $U(1)$ gauge coupling constant at the mediation
scale is too strong and has a Landau pole problem below the GUT scale, i.e. $g_X \gsim 0.5$.
In the light-shaded region, the radiatively generated potential does not curl up for the 
large value of $S$.
}
\label{fig:Rbreaking}
\end{figure}

Put the above discussions together,
we find that spontaneous R-symmetry breaking 
and the right amount of the entropy production are achieved 
at the same time for $\l = O(10^{-1})$ in the extended vector-like SUSY breaking model.

\section{Conclusion}
In this paper, we discussed
the simplest class of the R-invariant gauge mediation model 
for the gravitino with a mass in the one to ten keV range.
The gravitino dark matter scenario with the mass in this range 
is drawing attention as an interesting interpretation of 
seeming discrepancies 
between the observation and the simulation of the structure formation
based on the cold dark matter model.

For a consistent  gravitino dark matter scenario with a mass in the one to ten keV range,
the relic density of the gravitino is needed to be diluted by a factor of $\D = 10-100$.
In this paper, we  discussed entropy production from the 
vector-like SUSY breaking model, which is extended so that spontaneous 
R-symmetry breaking is achieved.
Spontaneous R-symmetry breaking is necessary ingredient for the successful
R-invariant gauge mediation mechanism.
As a result, we find that  R-symmetry breaking and right amount of entropy
production are achieved at the same time for a certain parameter space.

The interesting prediction of the R-invariant gauge mediation model is the 
peculiar gaugino mass spectrum with
much heavier sfermions.
Especially, the gluino can be lighter than 
the wino even if the messenger masses and coupling constants
satisfy the GUT relation at the GUT scale.
The light gluino ($m_{\rm gluino}\simeq 300\,{\rm GeV}-1$\,{TeV})
is quite advantageous to be produce at the LHC.
Therefore, it is expected that this model can be probed by the LHC.

Another notable point is that the solution to the $\mu$-problem proposed in Ref.~\cite{Yanagida:1997yf} 
works for the gravitino mass of order ${\cal O}(1)-{\cal O}(10)~\KEV$.
It is very interesting that the gravitino mass of this order is favored from several things, i.e.
the warm dark matter, the interesting region of the gaugino masses at the LHC,
and the solution to the $\mu$-problem.

\section*{Acknowledgements}
We would like to thank A.~Kamada, S.~Shirai and N.~Yoshida for useful discussions.
This work  was supported by the World Premier 
International Research Center Initiative (WPI Initiative), MEXT, Japan.
The work of RS and KY is supported in part by JSPS Research
Fellowships for Young Scientists.
\appendix
\section{More on the extended SUSY breaking sector}
\subsection{Mass spectrum}
\subsubsection*{Scalar spectrum}
In order to analyze the spectrum of the scalar particles, we decompose the scalar components
as,%
\footnote{In this paper, we assume that all the parameters in the
SUSY breaking sector are real valued, so that the $CP$-symmetry is not broken. However, notice that the $CP$ violation in the hidden sector is 
insignificant for the $CP$ violation of the SSM, because the phases of the gaugino masses are the same as $\vev{S}^*F$, which can always be rotated away.}
\begin{eqnarray}
 S  &=&
 \left( \vev{S} +  \frac{1}{\sqrt 2}\sigma \right) e^{i a/ {\sqrt 2\vev{S}}  }\ ,\quad
 T = \frac{1}{\sqrt 2}(x_T + i y_T) e^{i a/{\sqrt 2 \vev{S}}  } \ ,\cr
 S_a &=& \frac{1}{\sqrt 2}(x_s + i y_s) e^{ia/{\sqrt 2\vev{S}}}\ ,
 \quad
M_a = \frac{1}{\sqrt 2}(x_m + i y_m) \ , \cr
\phi &=& \frac{1}{\sqrt 2} (x_\phi + i y_\phi) \ .
\end{eqnarray}
In the followings, we suppress the index $a$ of $x_{s,m}$ and $y_{s,m}$.


The squared mass matrices of $x$'s are given by~\footnote{
Note that the low-energy effective K\"ahler potential of $\phi$ is given by $K\simeq \Lambda^2[(\phi+\phi^*)/\sqrt{2}\Lambda-2g_X V_X)]^2+\cdots$, where
$V_X$ is the gauge supermultiplet of $U(1)_X$.}
\begin{eqnarray}
{\cal M}^{(x)2}_{T\phi} = 
\left(
\begin{array}{ccc}
\l^2 \L^2 + 2 g_X^2 \vev S^2   & - \l^2 \L \vev S/{\sqrt 2}+2\sqrt{2} g_X^2 \L\vev{S} \\
- \l^2 \L {\vev S}/{\sqrt 2} +2\sqrt{2} g_X^2 \L\vev{S} &   
 2\l^2 \L^2 + \frac{1}{2}\l^2\vev{S}^2 +4 g_X^2 \L^2 
  \\
\end{array}
\right)\ ,
\end{eqnarray}
for $(x_T, x_\phi)$ and 
\begin{eqnarray}
{\cal M}^{(x)2}_{sm} = 
\left(
\begin{array}{ccc}
\l^{\prime 2} \L^2   & - \l\l^\prime \L \vev S/{\sqrt 2} \\
- \l \l^\prime \L {\vev S}/{\sqrt 2}   &   
(\l^{\prime 2}-\l^2) \L^2 + \frac{1}{2}\l^2 
 \vev{S}^2
  \\
\end{array}
\right)\ ,
\end{eqnarray}
for $(x_s, x_m)$.
On the other hand, the squared mass matrices of $y$'s are given by,
\begin{eqnarray}
{\cal M}^{(y)2}_{T\phi} = 
\left(
\begin{array}{ccc}
\l^2 \L^2   & - \l^2 \L \vev S/{\sqrt 2} \\
 -\l^2 \L {\vev S}/{\sqrt 2}   &   
  \frac{1}{2}\l^2 
 \vev{S}^2
  \\
\end{array}
\right)\ ,
\end{eqnarray}
for $(y_T, y_\phi)$ and 
\begin{eqnarray}
{\cal M}^{(y)2}_{sm} = 
\left(
\begin{array}{ccc}
\l^{\prime 2} \L^2   &  -\l\l^\prime \L \vev S/{\sqrt 2} \\
- \l \l^\prime \L {\vev S}/{\sqrt 2}   &   
(\l^{\prime 2}+\l^2) \L^2 + \frac{1}{2}\l^2 
 \vev{S}^2
  \\
\end{array}
\right)\ ,
\end{eqnarray}
for $(y_s,y_m)$.

The eigen-modes of each mass matrices are given by,
\begin{eqnarray}
m_1^2 &=& \frac{1}{2} \left( 
\tr {\cal M}^2 - \sqrt{ (\tr {\cal M}^2)^2 - 4 \det {\cal M}^2} 
\right)\ , \cr
m_2^2 &=& \frac{1}{2} \left( 
\tr{\cal M}^2 + \sqrt{ (\tr{\cal M}^2)^2 - 4 \det{\cal M}^2} 
\right)\  .
\end{eqnarray}
Notice that the lighter mode of $(y_T,y_\phi)$ is massless,  which corresponds
to the would-be Nambu-Goldstone mode of the gauged $U(1)$ symmetry.%
\footnote{There is no mixing between the R-axion and the would-be 
Nambu-Goldstone boson.  }

\subsubsection*{Fermion mass spectrum}
The mass matrix of the fermion components of $S_a$ and $M_a$ is given by
\begin{eqnarray}
\label{eq:massmatrixF1}
{\cal M}^{(f)}_{sm} = 
\left(
\begin{array}{cc}
0    &  \l' \L  \\
\l' \L    &   -\l \vev S /\sqrt 2
\end{array}
\right)\ .
\end{eqnarray}
The mass matrix of the fermion components of $T$, $\phi$ and the $U(1)$ gaugino is given by,
\begin{eqnarray}
{\cal M}^{(f)}_{T\phi\tilde g} =
\left(
\begin{array}{ccc}
0  & -\l \L  & -i \sqrt 2 g_X \vev S   \\
-\l \L  & \l \vev S/\sqrt{2}   & -i 2 g_X \L  \\
-i \sqrt{2} g_X \vev S  & -i 2 g_X\L  &   0
\end{array}
\right)\ .
\end{eqnarray}
The fermion component of $S$ corresponds to the Goldstino.

\subsubsection*{Vector boson mass}
The mass of the vector boson is given by,
\begin{eqnarray}
 M_V^2 = 2 g_X^2 ( \vev{S}^2 + 2 \L^2)\ .
 \end{eqnarray}
 
 \subsubsection*{The lightest particle}
As we have discussed, the model possesses a discrete $Z_4$ symmetry which is effectively
a $Z_2$ symmetry under which $x_{m,s}$ and $y_{m,s}$ are odd.
From the above mass matrices, we find that the lightest, and hence stable, $Z_2$ odd particle resides 
in $x_{m,s}$.
In the followings, we name the lightest particle $x_L$.

In addition to the lightest $Z_2$ odd particle, the model possesses 
two massless scalars which correspond to 
the pseudo-flat direction, $\s$, and the R-axion, $a$, respectively.
Besides, the model also has a massless fermion, the fermion component of $S$,
which corresponds to the Goldstino.
As we will discuss, the mass of the pseudo-flat direction is generated
by the radiative corrections.
The R-axion mass is also generated by the effects of the explicit R-symmetry breaking terms, e.g. the constant term in the superpotential generates the
axion mass in supergravity.
The Goldstino becomes massive by the super-Higgs mechanism of supergravity.

 \subsubsection*{Relevant interactions}
 The relevant interaction terms to analyze the relic density of
 the lightest $Z_2$ charged particle is summarized below.
 
First, the R-axion interactions only appear in the kinetic terms.
In the basis we have defined above, 
the R-axion interactions come from the kinetic terms of $S$, $T$ and $S_{a}$,
\begin{eqnarray}
\label{eq:kin}
{ \cal L } = \frac{1}{2} (\partial a)^{2}\left( 1 + \frac{\s}{\sqrt 2\vev S} \right)^{2}
+ \sum_{i =s,T}
 \frac{(\partial a)^{2}}{4 \vev S^{2} } 
 (x_{i}^{2}+y_{i}^{2} )
+ \frac{\partial_\m a }{\sqrt 2 \vev S} (x_{i}\partial^{\m} y_{i}  - y_{i}\partial^{\m} x_{i}).
\end{eqnarray}

The other important interaction term is 
\begin{eqnarray}
 {\cal L} = \frac{\l}{2} (x_m + i y_m){\tilde G} \psi_m
+  \frac{\l}{2} (x_m - i y_m){\tilde G}^\dagger \psi_m^\dagger \ , 
\end{eqnarray}
where $\tilde G$ denotes the Goldstino and $\psi_m$ the fermion component of $M_a$.

 \subsection{Annihilation of the lightest $Z_4$ charged ($Z_2$ odd) field}
The annihilation modes of $x_L$ are $x_Lx_L\to \tilde G\tilde G$,
$x_L x_L \to \sigma \sigma$, and $x_L x_L \to a a $.
In terms of $x_L$, the relevant interaction terms  can be rewritten by,
\begin{eqnarray}
 {\cal L}&=& \frac{\l^2\vev S}{2\sqrt{2}} \frac{m_L^2}{m_H^2-m_L^2} \sigma x_L^2 
 - \frac{\l^3\l^\prime}{4} \frac{\L(2 \L^2+\vev S^2)}{m_H^2-m_L^2}\sigma x_L x_H
 - \frac{1}{16}\frac{\l^4 \l^{\prime2} \L^2\vev S^2}{(m_H^2 - m_L^2)(m_H^2 - \l^{\prime 2}\L^2)}\sigma^2 x_L^2
\cr
&&+ \frac{\cos^2\theta_x}{4\vev{S}^2}(\partial a)^2 x_L^2 
+\frac{\cos\theta_x}{\sqrt{2}\vev S} \partial a
\left(
\cos\theta_y (x_L \partial y_L - y_L\partial x_L)
- \sin\theta_y (x_L \partial y_H - y_H\partial x_L)
\right)\cr
&&
+\frac{1}{2} (\partial a)^{2}\left( 1 + \frac{\s}{\sqrt 2\vev S} \right)^{2}
+ \frac{\l}{2}\sin\theta_x x_L \tilde G (\sin\theta_f \psi_L + \cos\theta_f \psi_H) + h.c.\ .
\end{eqnarray}
Here, $\theta_{x,y}$ are mixing angles of $x$ and $y$ components of $S_a$ and $M_a$,
and $\theta_f$ is the angle of the fermion components.
The mixing angles are given by,
\begin{eqnarray}
 \tan\theta_x = \frac{\sqrt{2}(m_L^2 - \l^{\prime 2} \L^2)}{-\l\l^\prime \L \vev S}\ ,
 \quad
  \tan\theta_y = \frac{\sqrt{2}(m_L^{(y)2} - \l^{\prime 2} \L^2)}{-\l\l^\prime \L \vev S}\ ,
  \quad
   \tan\theta_f = \frac{\sqrt{2}(m_L^{(f)2} - \l^{\prime 2} \L^2)}{-\l\l^\prime \L \vev S}\ .
\end{eqnarray}

\subsubsection{$x_Lx_L\to \sigma \sigma$}
The process, $x_Lx_L\to\sigma\sigma$, proceeds via the $t$ and $u$-channel exchanges of $x_L$
and $x_H$ as well as via the contact interaction.
The amplitude of the $t$ and $u$-channel $x_L$ exchange is  given by,
\begin{eqnarray}
 {\cal M}_{2x_L\to 2\sigma}&=&\frac{1}{2}\l^4 \vev S^2\left( \frac{m_L^2}{m_H^2-m_L^2} \right)^2 
 \left(\frac{1}{m_L^2-t}
 +\frac{1}{m_L^2-u}
 \right)\ ,\cr
 & \simeq &
\frac{1}{2} \l^2  \frac{\l^2\vev S^2m_L^2}
{ \left(m_H^2-m_L^2 \right)^2 }
 \ ,
\end{eqnarray}
where we have neglected the mass of the flaton and used $t\simeq u \simeq - m_L^2$ 
in the non-relativistic limit.
The amplitude
 of the $t$ and $u$-channel $x_H$ exchanges is  given by,
\begin{eqnarray}
 {\cal M}_{2x_L\to 2\sigma}&=&\frac{\l^6\l^{\prime 2}}{16}
 \L^2\left( \frac{2 \L^2 + \vev S^2}{m_H^2-m_L^2} \right)^2 
 \left(\frac{1}{m_H^2-t}
 +\frac{1}{m_H^2-u}
 \right)\ ,\cr
 & \simeq &
 \frac{\l^2}{8}
\left( \frac{2\l^2 \L^2 + \l^2\vev S^2}{m_H^2-m_L^2} \right)^2 
 \left(\frac{\l^{\prime 2} \L^2}{m_L^2+m_H^2}\right)
 \ .
\end{eqnarray}
The amplitude of the contact term interaction is given by,
\begin{eqnarray}
 {\cal M}_{2x_L\to 2\sigma}&=&-\frac{\l^2}{4}
 \left(
 \frac{\l^{2} \vev S^2
 }{m_H^2-m_L^2} \right)
 \left(
  \frac{\l^{\prime 2} \L^2
 }{m_H^2-\l^{\prime 2}\L^2}
\right) \ .
\end{eqnarray}
Notice that the sum of all the above matrix elements vanishes for $\l = \l^\prime$.

\subsubsection{ $x_Lx_L\to aa$}
The process $x_Lx_L\to aa$ has four contributions, from the contact term interaction, the $t$ and 
$u$-channel exchange of $y_L$ and the $s$-channel flaton exchange.
The matrix element of the contact term interaction is given by, 
\begin{eqnarray}
 {\cal M}_{x_Lx_L\to aa} 
 =- \cos^2\theta_x\frac{p_3\cdot p_4}{\vev S^2}\simeq -2\cos^2\theta_x \frac{m_L^2}{\vev{S}^2}\ ,
\end{eqnarray}
where we have neglected the mass of the R-axions in the final state.
The matrix element of the $t$-channel exchange of $y$'s is given by,
\begin{eqnarray}
{\cal M}_{x_Lx_L\to aa} &=& \frac{\cos\theta_x^2\cos^2\theta_y}{2 \vev S^2}
\left(\frac{p_3\cdot(-2p_1+p_3)\,p_4\cdot(-2p_2+p_4) }{m_L^{(y)2}-t}
+
\frac{p_4\cdot(-2p_1+p_4)\,p_3\cdot(-2p_2+p_3) }{m_L^{(y)2}-u}
\right) \cr
&\simeq& \frac{4\cos\theta_x^2\cos^2\theta_y}{\vev S^2}\frac{m_L^4}{m_L^2+m_L^{(y)2}}\ ,
\end{eqnarray}
where we again neglected the mass of the R-axions.
The contributions from the $y_H$ exchanges are given by,
\begin{eqnarray}
{\cal M}_{x_Lx_L\to aa}
\simeq \frac{4\cos^2\theta_x\sin^2\theta_y}{\vev S^2}\frac{m_L^4}{m_L^2+m_H^{(y)2}}\ .
\end{eqnarray}
Finally, the flaton exchange contribution is given by,
\begin{eqnarray}
{\cal M}_{x_L x_L\to aa} = \l^2 \frac{m_L^2}{m_H^2-m_L^2}\frac{p_3\cdot p_4}{s}\simeq
\frac{\l^2}{2} \frac{m_L^2}{m_H^2-m_L^2}\ .
\end{eqnarray}

\subsubsection{$x_Lx_L\to \tilde G \tilde G$}
The annihilation cross section into the gravitino is given by $t$ and $u$-channel 
exchange of $M_a$ and $S_a$ fermions.%
\footnote{
Here, we neglect the $s$-channel flaton decay which utilizes
the higher dimensional operators.
}
The matrix element is given by,
\begin{eqnarray}
 {\cal M}_{x_L x_L \to \tilde G\tilde G} = \frac{\l^2}{4}\sin^2\theta_x\sin^2\theta_f
\, \left(\bar{u}(p_3)\frac{\slashchar q_t + m_L^{(f)} }{m_L^{(f)2}-t}v(p_4)
+
\bar{u}(p_3)\frac{\slashchar q_u + m_L^{(f)} }{m_L^{(f)2}-u}v(p_4)
\right)\ ,
 \end{eqnarray}
where $q_t = p_3-p_1$ and $q_u = p_3-p_2 $.
The heavier field exchange is then given by,
\begin{eqnarray}
 {\cal M}_{x_L x_L \to \tilde G\tilde G} = \frac{\l^2}{4}\sin^2\theta_x\cos^2\theta_f
\, \left(\bar{u}(p_3)\frac{\slashchar q_t + m_H^{(f)} }{m_H^{(f)2}-t}v(p_4)
+
\bar{u}(p_3)\frac{\slashchar q_u + m_H^{(f)} }{m_H^{(f)2}-u}v(p_4)
\right)\ ,
 \end{eqnarray}

Thus, the unpolarized squared amplitude is given by,
\begin{eqnarray}
|{\cal M}_{x_Lx_L\to \tilde G\tilde G}|^2 
&\simeq&  2\l^4  \sin^4\theta_x\sin^4\theta_f\frac{m_L^{(f)2}m_L^2}{(m_L^2+m_L^{(f)2})^2}
+ 2\l^4  \sin^4\theta_x\cos^4\theta_f\frac{m_H^{(f)2}m_L^2}{(m_L^2+m_H^{(f)2})^2}\cr
&&- 4\l^4  \sin^4\theta_x\sin^2\theta_f\cos^2\theta_f
\frac{|m_L^{(f)}m_H^{(f)}|m_L^2}{(m_L^2+m_L^{(f)2})^2}\ .
\end{eqnarray}
Here, we have used the fact that the product of the two fermion masses
is the negative valued since the determinant of the mass matrix in Eq.\,(\ref{eq:massmatrixF1}) is negative.

\subsubsection{Total cross section}
By adding up all the above modes, we obtain the total $S$-wave cross section,
\begin{eqnarray}
\sigma v_{\rm rel}  \simeq \frac{1}{2}\frac{1}{32\pi}\frac{|{\cal M}|^2}{m_L^2}\ ,
\end{eqnarray}
where a factor $1/2$ represents the statistical factor of the final state particles.
The thermal average can be trivially taken and the resultant cross section appearing 
in the Boltzmann equation is given by,
\begin{eqnarray}
\vev{ \sigma v_{\rm rel} } \simeq\frac{1}{2} \frac{1}{32\pi}\frac{|{\cal M}|^2}{m_L^2}\ .
\end{eqnarray}

\subsection{Decay of the $Z_2$ charged particle}
As we have discussed in the paper, 
the $Z_2$ symmetry is expected to be broken by the reduced Planck suppressed 
operators in Eq.\,(\ref{eq:Z2b}).
This operator generates the interaction terms of the $Z_2$ odd particle such as,
\begin{eqnarray}
 {\cal L} &=&\left(\frac{\l \vev{S} M_a}{\sqrt{2}}-\l' \L S_a \right) \frac{c \L}{M_{P}} M_a (H_u H_d)^* - \frac{c\L}{M_P}M_a \psi_{H_u} \psi_{H_d} + h.c. \cr
& \to&  \left( \frac{1}{2} \sin\theta_x\l \vev{S} - \frac{1}{\sqrt{2}}\cos \theta_x \l' \L \right) \frac{c\L}{M_{P}} x_L(H_u H_d)^* 
-\sin\theta_x   \frac{c\L}{\sqrt 2 M_P}x_L \psi_{H_u}\psi_{H_d} + h.c. \nonumber \\
\end{eqnarray}
Thus, the decay rate of the lightest $Z_2$ odd particle is given by,
\begin{eqnarray}
\G_{x_L} \simeq \frac{1}{4\pi}\left( \frac{1}{2} \sin\theta_x\l \frac{\vev{S}}{m_{x_L}} - \frac{1}{\sqrt{2}}\cos \theta_x \l' \frac{\L}{m_{x_L}} \right)^2
\frac{c^2 \L^2}{M_{P}^2}m_{x_L}
+ \frac{1}{8\pi}\sin^2\theta_x \frac{c^2 \L^2}{M_{P}^2}m_{x_L}\ .
\end{eqnarray}

\subsection{The messenger interaction}
For the extended vector-like SUSY breaking model, we need at least two sets of the messengers, which couple to $S_\pm$, 
\begin{eqnarray}
\label{eq:messenger2}
W &=& \left( \tilde{\Psi}_1, \tilde{\Psi}_{1+} \right) \left(
\begin{array}{cc}
k^{(-)}S_+ & m \\
m & 0 
\end{array}
\right)
\left(
\begin{array}{c}
\Psi_{1-} \\
 \Psi_1
\end{array}
\right)
+\left( \tilde{\Psi}_2, \tilde{\Psi}_{2-} \right) \left(
\begin{array}{cc}
k^{(+)}S_- & m \\
m & 0 
\end{array}
\right)
\left(
\begin{array}{c}
\Psi_{2+} \\
 \Psi_2
\end{array}
\right) \nonumber \\
&\simeq&  \left( \tilde{\Psi}_1, \tilde{\Psi}_{1+} \right) \left(
\begin{array}{cc}
kS & m \\
m & 0 
\end{array}
\right)
\left(
\begin{array}{c}
\Psi_{1-} \\
 \Psi_1
\end{array}
\right)
+\left( \tilde{\Psi}_2, \tilde{\Psi}_{2-} \right) \left(
\begin{array}{cc}
kS & m \\
m & 0 
\end{array}
\right)
\left(
\begin{array}{c}
\Psi_{2+} \\
 \Psi_2
\end{array} \right) \ ,
\end{eqnarray}
where we have rewritten $S_{\pm}$ in terms of the pseudo-flat direction $S$, and we have taken $k_{(+)}=k_{(-)}=\sqrt{2}k$.

Thus, the Dirac-type fermion mass matrix is given by,
\begin{eqnarray}
 {\cal M}^{(f)}=
\left(
\begin{array}{cc}
 k \vev{S} & m  \\
   m & 0
\end{array}
\right)\ .
\end{eqnarray}
for each $\Psi_1$'s and $\Psi_2$'s.
The squared mass matrix of the complex scalars is given by,
\begin{eqnarray}
{\cal M}^{(s)}=
\left(
\begin{array}{cccc}
{k^2}\vev{S}^2 + m^2  &  k m \vev{S}  & k F   &0\\
 k m \vev{S}  & m^2  &  0 &0\\
k F  &  0 & {k^2}\vev{S}^2 + m^2    & k m \vev{S} \\
0   & 0  & k m \vev{S}   &m^2
\end{array}
\right)\ .
\end{eqnarray}

The messenger sector has the $U(1)_{d1}\times U(1)_{\ell1} \times U(1)_{d2} \times U(1)_{\ell2}$ global symmetries, which act on
the down type and lepton type $\Psi_1$'s and $\Psi_2$'s, respectively. We have to break these symmetries explicitly to 
avoid the messenger dark matter overclosing the universe. 
For example, by introducing a small mixing such as $\d m \tilde{\Psi}_1 \Psi_2$, we can break the symmetry down to
$U(1)_{d}\times U(1)_{\ell}$. Furthermore, we can completely break the symmetry
by introducing interactions such as $\e \tilde{\Psi}_1 {\bf 10}_{\rm SSM}{\bf 5}^*_{\rm SSM}$, where ${\bf 10}_{\rm SSM}$ and ${\bf 5}^*_{\rm SSM}$
are the SSM matter fields written in the $SU(5)_{\rm GUT}$ representations and $\e$ is a very small Yukawa coupling.
What type of interactions are allowed depends on the precise R-charge assignment to the SSM matter fields, which we do not explicitly specify in this paper.

\end{document}